# Surface segregation of liquid metal plasma-facing component alloys: A ReaxFF investigation

Md Adnan Mahathir Munshi,[1] Abdul Aziz Shuvo,[1] Mike Kotschenreuther,[2] Adri C.T. van Duin,[1] and Bladimir Ramos-Alvarado[1*]

[1]Department of Mechanical Engineering, The Pennsylvania State University, University Park, PA-16802, USA.

[2]ExoFusion, Bellevue, WA-98005, USA.

**Abstract**

Engineering liquid metal alloys offers a transformative pathway for plasma-facing components (PFC) by enabling chemically tailored surfaces that can simultaneously optimize plasma-material interactions, reduce divertor heat flux, and enhance core plasma confinement, thereby advancing the commercial viability of nuclear fusion power plants. This study, employing an atomistic simulation framework, provides direct evidence that incorporating non-metal surface-active agents (such as O and H, or their combination) enables strong surface segregation. This capability makes tin–aluminum (Sn–Al) and tin-lithium (Sn–Li) alloys, with suitable compositions, good candidates for PFC applications. Specifically, the presence of low-Z solutes (Li, Al) leads to preferential surface enrichment, which imparts low-Z sputtering characteristics, while the Sn solvent maintains thermophysical stability. To systematically examine this behavior, we first optimized ReaxFF parameter sets for Sn–Al, Sn–Al–O, Sn–Li, Sn–Li–O, Sn–Li–H, and Sn–Li–O–H systems. We validated them using formation energies and elastic constants. We then employed reactive molecular dynamics simulations to resolve the coupled effects of surface segregation and impurity-driven chemistry at fusion-relevant temperatures. We also introduced an overlap-based segregation index that captures interfacial compositional separation directly from atomistic density distributions. This metric reveals a clear hierarchy of segregation regimes across all systems and presents a unified view of segregation across all observations reported herein. Together, these findings establish a mechanistic link between non-metal chemistry and interfacial structure, providing a predictive framework for designing self-adaptive, low-sputtering liquid metal alloys for fusion applications.

*Corresponding Author: Email Address: bzr52@psu.edu

## 1. Introduction

Liquid metals (LM) have emerged as leading candidates for plasma-facing components in magnetic confinement fusion devices, serving as the primary interface between the hot fusion plasma and the reactor wall.[1,2] Unlike conventional solid PFCs, LM PFCs offer intrinsically self-healing surfaces and enhanced tolerance to extreme thermal loads-from steady-state heat fluxes exceeding 10–20 MW $m^{-2}$,[3] to transient events, such as edge-localized modes (ELMs) and disruptions, that can exceed 0.5–1 GW $m^{-2}$ on sub-millisecond timescales.[4,5] These benefits are observed while also mitigating plasma-driven dust formation and enabling radiative power dissipation via controlled evaporation,[4] all without permanent structural failure.[6,7] These attributes make them attractive for applications spanning tokamaks, spherical tokamaks, stellarators, and field-reversed configurations across both confining and non-confining magnetic geometries,[8] thereby positioning them as promising alternatives to solid tungsten. [5] However, achieving fusion power plant economic viability requires plasma confinement efficiency (H-factor) above 1.5. [9] Tungsten, the current baseline solid PFC material, achieves $H < 1.2$ at reactor-relevant densities compared to carbon PFCs' historical H = 1.5-1.8 (and recent 2024 DIII-D results reaching H = 2.5).[10–14] In contrast, liquid lithium, while avoiding high-Z radiation losses, exhibits equally severe limitations due to poor divertor radiative efficiency. SOLPS modeling of STEP geometry shows that Li-based systems fail to reach the 10 MW $m^{-2}$ engineering limit even at separatrix densities of $11 \times 10^{19}$ $m^{-3}$ (versus $3.5 \times 10^{19}$ $m^{-3}$ for Argon seeding), inherently suppressing density peaking and degrading H-factor. As neither tungsten nor pure LMs can meet the coupled demands of heat exhaust and confinement; consequently, engineered LM alloys could address this by reducing high-Z radiation while preserving core-edge plasma integration. [4] Recently, the concept of greatly improving PFC performance by chemically engineered surface segregation has been proposed.[15] That concept is investigated here.

The performance of binary alloys as PFCs is governed by a fundamental design principle: combining the advantages of low- and high-Z elements. Crucially, sputtering is governed by the atomic monolayer at the surface; thus, surface segregation directly determines the species interacting with the plasma.[16] Experimental evidence demonstrates that even trace elements can dominate surface behavior. For instance, in the Ga-Bi system, Bi constitutes only ~0.2% of the bulk yet accounts for ~80% of the sputtered flux due to strong surface segregation,[17] consistent with classical Gibbs segregation driven by surface tension differences. This principle enables a high-Z solvent to provide low vapor pressure and thermal stability, while a low-Z solute dominates the plasma-facing surface. Historically, segregation was modeled using the Butler equations,[18] which attribute enrichment to the energy difference between surface and bulk atoms, classically simplified as a function of the surface tension (σ):

$$S \sim e^{\frac{(\sigma_a - \sigma_b)A}{kT}}$$

However, the associated driving force (~0.1 eV per atom) is too weak to explain the experimentally observed near-complete surface enrichment. For instance, in Sn–Li alloys, recent thermodynamic studies[19] and ab initio simulations[20] show that surface tension effects alone predict only moderate lithium enrichment (S~0.35-0.4), insufficient to explain experimental sputtering data showing surfaces composed of 99% Li. [21] An explanation for this “sputtering paradox” lies in non-metals driven chemical segregation. While surface tension provides a weak driving force, chemical bond energies are in the range of several eV per atom, an order of magnitude increase in the exponent of the segregation factor. Because Li binds significantly more strongly to oxygen than Sn does, trace non-metals (even at concentrations 1-10 ppm) introduce a powerful chemical potential gradient. [22] This gradient induces preferential migration of low-Z elements to the surface to bond with oxygen. Time-resolved experiments by Bastaz [23] further confirm that upon melting, Li and O rapidly dominate the interface while Sn is depleted, with reappearance only after oxygen outgassing. Consequently, chemically driven segregation enables near-complete low-Z surface coverage, effectively decoupling surface and bulk properties, thereby shielding the high-Z bulk while maintaining favorable plasma-facing behavior.

Tailoring chemical gradients through non-metal incorporation offers a powerful pathway to overcome the intrinsic limitations of binary alloys in PFC design. Among binary alloys, tin–lithium (Sn–Li) alloys have been widely explored, typically with Li concentrations of 20–30 at%, where Li preferentially segregates to the surface. Extending this concept, multicomponent alloy design can be achieved by combining low-melting, low-vapor-pressure liquid metals such as Sn, Ga, or In with higher-melting, low-Z elements (e.g., Al to enhance thermal conductivity, Ca, or Li), while maintaining thermophysical properties compatible with fusion-relevant conditions. The addition of surface-active species such as O, H, or their combination further strengthens segregation by creating a chemical potential gradient that energetically favors low-Z elements at the surface. Accordingly, in this study, Sn–Al, Sn–Al–O, Sn–Li, Sn–Li–O, Sn–Li–H, and Sn–Li–O–H systems are investigated as promising candidates, where O/H-driven segregation decouples plasma-facing behavior from bulk composition while preserving the stability of the Sn matrix. Molecular dynamics (MD) is a powerful tool for capturing the coupled effects of surface segregation, impurity-driven chemistry, and dynamic bond rearrangement; however, it requires a modeling framework beyond conventional non-reactive force fields.[24] The ReaxFF reactive force field, [25–27] based on a bond-order formalism and trained on quantum-mechanical data, enables dynamic bond formation and dissociation, charge redistribution, and resolution of transient chemical species under finite-temperature conditions. Its capability to model complex multicomponent systems has been demonstrated in prior studies, [28] including Fe–Al–Ni alloys, where it successfully captured diffusion and segregation behavior. Moreover, ReaxFF enables relatively large-scale, long-timescale simulations with near *ab initio* fidelity, bridging atomic-scale chemistry and macroscopic PFC performance.[24] Despite these advances, atomic-scale mechanisms of non-metal-mediated segregation and interfacial stabilization in liquid metal alloys remain poorly understood, limiting the design of robust next-generation PFC materials with enhanced resistance to sputtering, evaporation, and plasma-induced degradation, thereby motivating this atomistic investigation.

Leveraging this ReaxFF-based molecular dynamics framework, we investigated the atomic-scale mechanisms governing intrinsic and non-metal (O, H)-mediated surface segregation in Sn–Al and Sn–Li alloys under fusion-relevant conditions. The force fields were systematically developed and validated against DFT formation energies, convex-hull thermodynamics, and elastic constants to ensure reliable reactive behavior. A physically consistent workflow was implemented in which alloy slabs were equilibrated, exposed to O/H radicals, and subsequently heated to promote diffusion-driven segregation. The resulting structures were analyzed using probability density functions, local composition profiles, and atomistic visualization, alongside a unified segregation metric enabling direct comparison across systems. This integrated approach establishes a mechanistic link between non-metal-driven surface chemistry and segregation behavior in multicomponent liquid metal PFC alloys, providing atomistic insights for the rational design of next-generation plasma-facing materials capable of simultaneously meeting the stringent heat exhaust and plasma confinement requirements necessary for economically viable fusion power plants.

## 2. Methodology

### 2.1 Topology of nanostructures

To capture surface-driven phenomena relevant to plasma-facing conditions, all systems were constructed using a slab-based nanostructure with two free surfaces. As illustrated in Figure 1(a–f), the initial models were generated from a diamond-cubic Sn lattice containing 2000 atoms with a lattice constant of 6.5 Å using the Atomsk tool.[29–31] The diamond-cubic lattice provides a neutral and computationally convenient starting configuration for doping with Al or Li; subsequent equilibration and high-temperature ramping eliminate structural memory, yielding a physically representative liquid state independent of the starting lattice. Alloy compositions were obtained via random substitution of Sn atoms with Al or Li, producing chemically homogeneous structures. For the Sn–Al system, a composition of 7.4 wt% Al was generated by substituting the appropriate number of Sn atoms with Al. For the Sn–Li systems, Li was introduced at 9 at%, 20 at%, and 26 at%, corresponding to 180, 400, and 520 Li atoms (≈2 wt%), respectively. In this study, the 26 at% (≈2 wt%) composition is primarily emphasized, as it enables direct comparison with prior *ab initio* molecular dynamics investigations. The 20 at% composition is included for benchmarking against experimental studies, while the lower concentration of 9 at% is explored as a novel compositional regime relevant to plasma-facing component (PFC) applications.

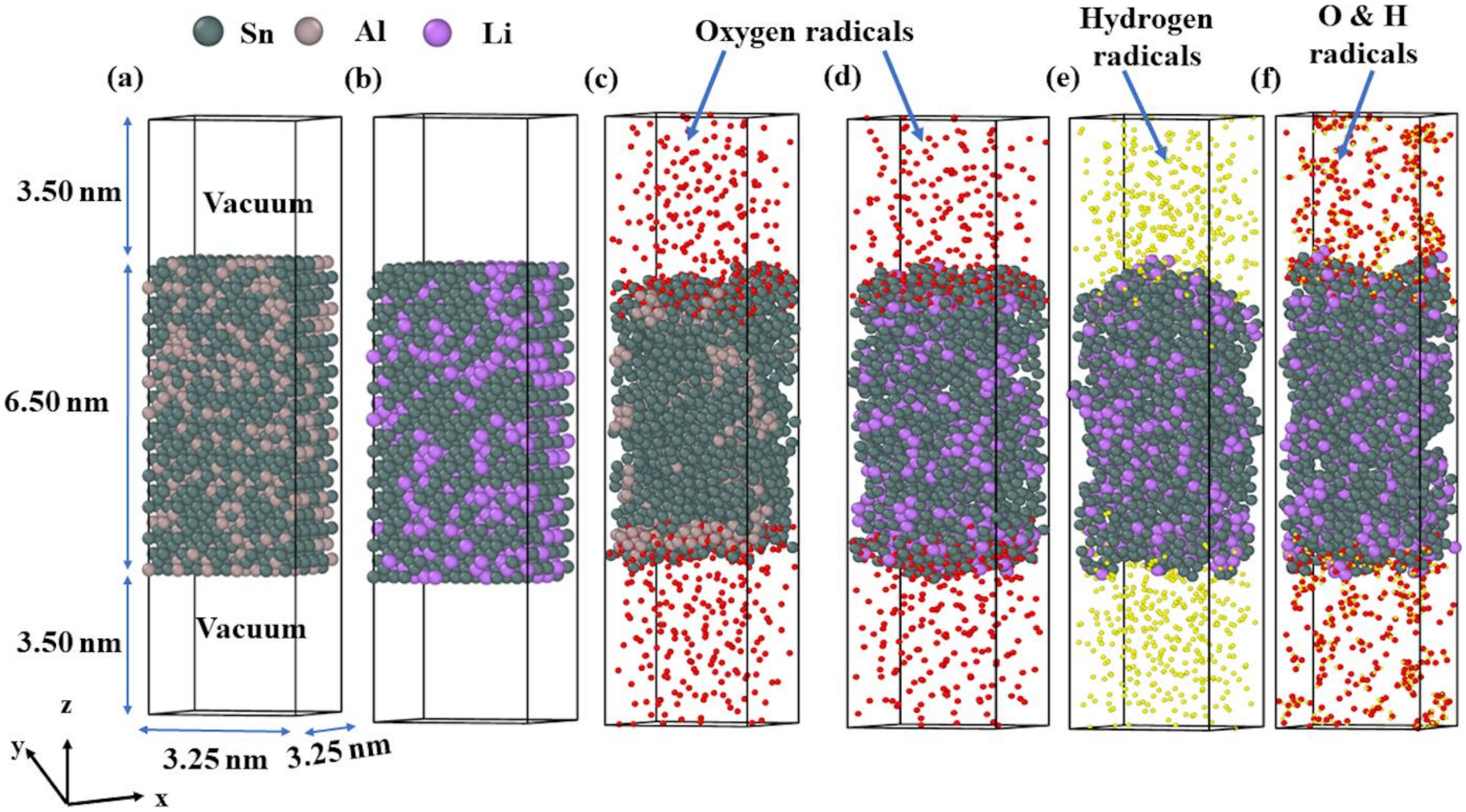

Figure 1. Initial configurations of Sn–Al and Sn–Li alloy slabs and their exposure to oxygen, hydrogen and combination of both radicals. (a, b) Sn–Al and Sn–Li slabs, each 6.5 nm thick, are positioned between 3.5 nm vacuum layers along the z-axis. (c, d, e, and f) After equilibration at 300 K, the systems are exposed to oxygen radicals (red), hydrogen radicals (yellow), and both oxygen and hydrogen (red and yellow), which are introduced above and below the slabs. Sn, Al, and Li atoms are shown in teal, brown, and purple, respectively. The coordinate axes denote the orientation of the simulation box.

The resulting alloy structure was constructed as a finite slab with in-plane dimensions of 3.25 nm × 3.25 nm and a thickness of 6.5 nm. Periodic boundary conditions were applied in all three directions, while a vacuum region of 3.5 nm was introduced on both sides along the z-direction to effectively generate two free surfaces (Figure 1a, b). This topology represents a two-dimensional periodic slab with symmetric dual interfaces, allowing simultaneous analysis of segregation at both surfaces while avoiding artificial interactions across periodic boundaries. Following equilibration at 300 K, reactive environments were introduced by inserting oxygen and/or hydrogen radicals into the vacuum regions. These species were symmetrically distributed above and below the slab, ensuring identical chemical exposure at both interfaces. The placement of radicals was carried out to control spatial distribution and prevent initial atomic overlap. As shown in Figure 1(c–f), this procedure yields a series of progressively complex interfacial environments, including oxygen-only, hydrogen-only, and combined O–H systems, enabling direct comparison of their individual and synergistic effects on segregation behavior.

A comprehensive set of simulation systems was constructed to systematically isolate the roles of alloy composition and reactive species, as summarized in Table 1. The study includes baseline binary alloys (Sn–Al and Sn–Li) without reactive species, as well as reactive configurations incorporating oxygen, hydrogen, and combined O–H environments. The number of introduced radicals was carefully scaled with alloy composition (e.g., 216–504 O radicals and 126 H radicals), ensuring consistent surface coverage and enabling meaningful comparisons across systems. These configurations allow the investigation of oxygen-induced, hydrogen-mediated, and synergistic segregation mechanisms under controlled conditions.

Table 1: Summary of simulation systems and compositions

| System Type | Composition (solute) | Reactive Species | Role in Study |
|---|---|---|---|
| Sn–Al (wt%) | 7.4% Al | None | Baseline binary alloys |
| Sn–Al–O | 7.4% Al | 216 O radicals | Oxygen-induced segregation analysis |
| Sn–Li (at%) | 9% and 26% Li | None | Baseline binary alloys |
| Sn–Li–O | 9% and 26% Li | 504 O radicals | Oxygen-mediated surface segregation |
| Sn–Li–H | 9% and 26% Li | 126 H radicals | Hydrogen-mediated surface segregation |
| Sn–Li–O–H | 9%, 20% and 26% Li | 288 O + 216 H | Synergistic reactive environment effects |

Overall, the resulting nanostructure consists of three distinct regions: (i) a bulk-like interior representing the alloy matrix, (ii) two interfacial regions where segregation and chemical interactions occur, and (iii) surrounding vacuum regions containing reactive species. This double-interface slab topology provides a physically consistent and computationally efficient framework for capturing surface segregation, oxidation, hydrogen interaction, and coupled O–H effects under high-temperature plasma-facing conditions, while enhancing statistical reliability through simultaneous sampling of two equivalent surfaces.

### 2.2 Development and validation of a ReaxFF force field for Sn/Al/Li/H/O interactions.

ReaxFF force field training was performed using the standalone ReaxFF program, using a single-parameter search algorithm, as described in more detail by van Duin et al.[32] DFT data used in the training were obtained from the online Materials Project database.[33] The parameterization is fully detailed in the Supplementary Material.

### 2.3 Molecular dynamics simulation protocol

The molecular dynamics workflow implemented in this study is illustrated in Figure 2. Following the slab construction (Figure 1a and 1b), equilibration of the Sn–Al and Sn–Li alloy systems was carried out using the newly developed Sn/Al/Li/O/H reactive force field within the LAMMPS simulation package.[34] Periodic boundary conditions were imposed in all spatial dimensions. Atomic velocities were initialized using a Gaussian distribution corresponding to a temperature of 300 K. To ensure statistical robustness and assess the sensitivity of the system to initial conditions, multiple independent simulations were performed using different random seed values for velocity initialization. Before dynamical simulation, the systems underwent conjugate-gradient energy minimization to eliminate unfavorable atomic overlaps and achieve mechanically stable configurations. At each timestep, charge equilibration was executed using the charge equilibration (QEq) method, capturing the dynamic charge redistribution intrinsic to the ReaxFF framework. A timestep of 0.25 fs was employed throughout all stages of the simulation to ensure numerical stability.

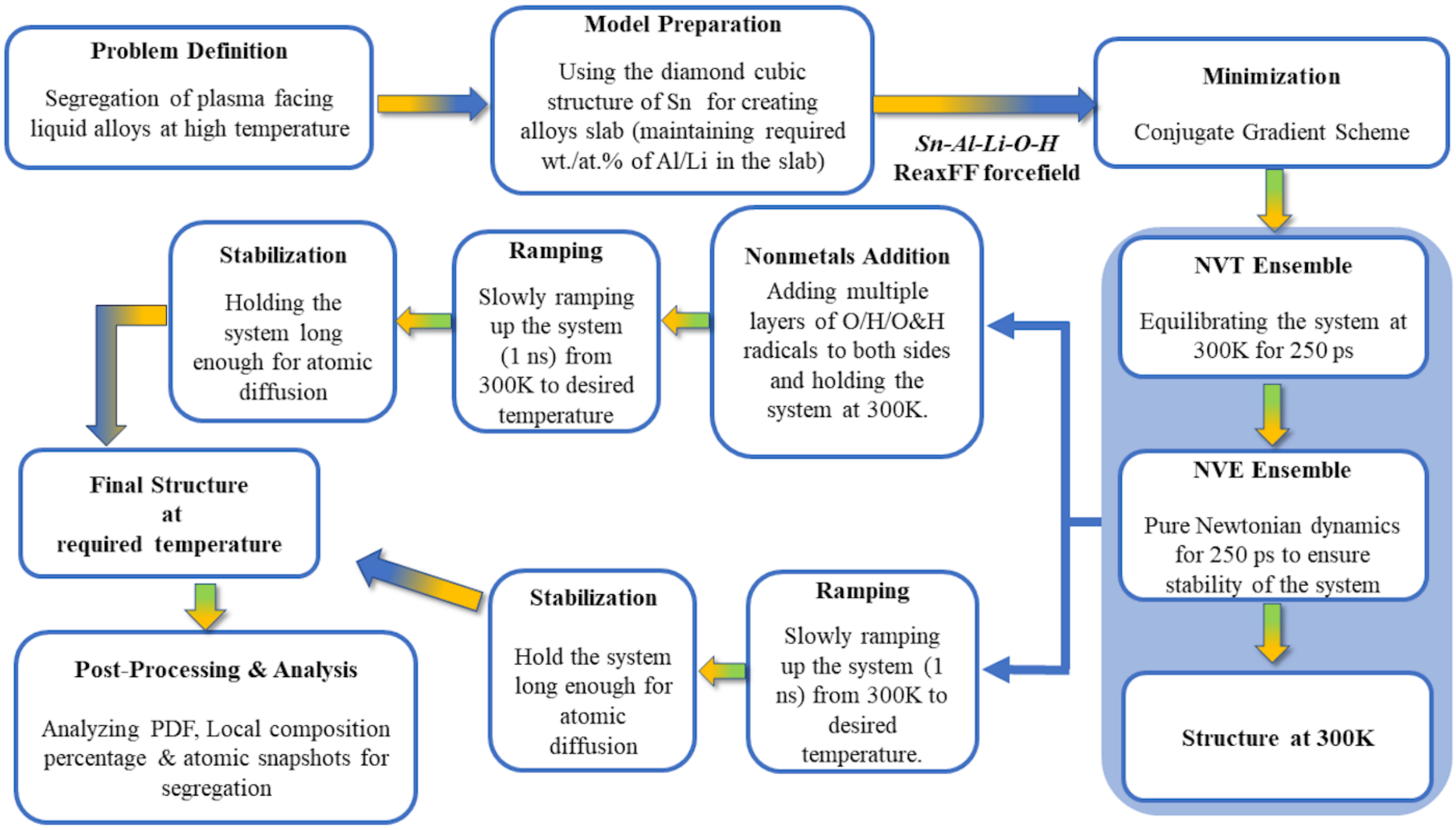

Figure 2. Schematic workflow of the ReaxFF molecular dynamics simulations used to investigate segregation in Sn–Al and Sn–Li plasma-facing liquid alloys.

(i) Following initialization, the system was equilibrated in the canonical (NVT) ensemble at 300 K using a Nose–Hoover thermostat with a damping parameter of 100 fs. To prevent artificial drift of the center of mass and preserve in-plane dynamics, a recentering procedure was applied along the z-direction throughout all thermostat-controlled stages. NVT equilibration lasted 250 ps to ensure attainment of thermal equilibrium. Afterward, both the thermostat and recentering procedure were removed, and the system was propagated in the microcanonical (NVE) ensemble for 250 ps. This second phase verified robust energy conservation and excluded artifacts potentially arising from the thermostat.

(ii) For the investigation of high-temperature segregation behavior of the Sn–Al and Sn–Li alloys, the equilibrated configurations from (i) were heated from 300 to 1300 K in the NVT ensemble using a Nose–Hoover thermostat with a damping parameter of 100 fs. The temperature ramp was performed over 1 ns, enabling controlled heating and structural relaxation without introducing thermal shocks. After reaching 1300 K, the system was further equilibrated in the NVT ensemble for an additional 0.5 ns to ensure thermal and structural stability at high temperature. Subsequently, both the thermostat and recentering constraints were removed, and the system evolved in the NVE ensemble for 250 ps.

(iii) To characterize high-temperature non-metal-induced segregation behavior in the Sn–Al and Sn–Li alloy systems, the atomic velocities were reassigned for structures containing oxygen, hydrogen, and combined hydrogen–oxygen radicals using different random seed values, as specified above, according to a Gaussian distribution corresponding to a temperature of 300 K, ensuring a physically consistent initial state while removing any net linear momentum. The systems were subsequently re-equilibrated under the NVT ensemble as in step (i). The systems were then heated from 300 to 1300 K in the NVT ensemble as in step (ii). For the Sn–Li system in the presence of both oxygen and hydrogen, additional simulations were performed at intermediate target temperatures of 800 K and 1000 K, alongside 1300 K, to enable benchmarking of segregation behavior across a representative high-temperature range. Throughout the thermostat-controlled stages, a recentering operation was maintained to prevent artificial drift of the slab. Subsequently, both the thermostat and recentering constraints were removed, and the systems continued in the NVE ensemble for 250 ps.

Throughout the final NVE phases, the atomic trajectories were sampled at regular intervals of 250 fs for subsequent analysis. One-dimensional probability density functions (PDFs) were computed by discretizing the simulation domain along the surface-normal ($z$) direction into uniform bins of 2 Å and time-averaging the atomic number density within each bin over the full 250 ps NVE production run, corresponding to 1000 sampled snapshots. In addition, local atomic composition profiles were obtained by evaluating the time-averaged atomic percentages of each species within the same spatial bins, along with the total atom count distribution. Furthermore, radial distribution function (RDF) calculations were performed, and atomic configurations were visualized using OVITO[35] to provide complementary structural insights into local ordering, interfacial structure, and species distribution.

To quantitatively characterize segregation behavior, an overlap-based metric was defined using the discretized number density profiles of the primary alloy species, as given in Equation (1). This formulation evaluates the degree of spatial co-localization between species by computing a normalized overlap of their density distributions along the surface-normal direction. The continuous integral form of this metric was numerically approximated using bin-wise summation consistent with the discretized nature of the simulation data. The corresponding segregation strength was then defined as *1-S*, where lower overlap indicates stronger spatial separation between species.

$$\boldsymbol{S} = \frac{\sum_i \boldsymbol{\rho_1}(\boldsymbol{z_i})\boldsymbol{\rho_2}(\boldsymbol{z_i})\boldsymbol{\Delta z}}{\sqrt{[\sum_i \boldsymbol{\rho_1^2}(\boldsymbol{z_i})\boldsymbol{\Delta z}][\sum_i \boldsymbol{\rho_2^2}(\boldsymbol{z_i})\boldsymbol{\Delta z}]}} \quad (\mathbf{1})$$

These post-processing analyses enabled quantifying the segregation behavior and surface enrichment, allowing examination of high-temperature, fusion-relevant segregation phenomena and elucidation of the role of non-metallic species, particularly oxygen and hydrogen, in driving alloy segregation and interfacial compositional evolution.

# 3. Results and Discussion

## 3.1 Intrinsic and oxygen-driven surface segregation in liquid Sn–Al alloys

Figures 3(a), 3(b), and 3(c) collectively elucidate the evolution of atomic-scale mixing, segregation, and interfacial chemistry in Sn–Al and Sn–Al–O slab systems under different thermodynamic conditions. The results reveal a clear progression from a nearly homogeneous alloy to a strongly segregated, oxygen-stabilized interfacial structure. Figure 3(a) represents a reference state of the Sn–Al system, where the two species remain largely well mixed. The probability density functions (PDFs) of Sn and Al, obtained from three independent simulations with different initial velocity seeds (Seeds 1–3), exhibit significant overlap across the slab-normal direction ($z$), indicating that both species are distributed throughout the system without pronounced interfacial or surface enrichment. While the Al profiles show slight variations in peak positions and amplitudes among different seeds, the overall trends remain consistent, demonstrating that the observed behavior is not sensitive to the initial velocity distribution. This behavior is further supported by the corresponding atomistic snapshot, where Al atoms are dispersed within the Sn matrix without forming a continuous segregated layer. Overall, these results indicate that, under the given conditions, the Sn–Al alloy remains predominantly mixed, with only weak compositional fluctuations and no strong thermodynamic driving force for large-scale segregation.

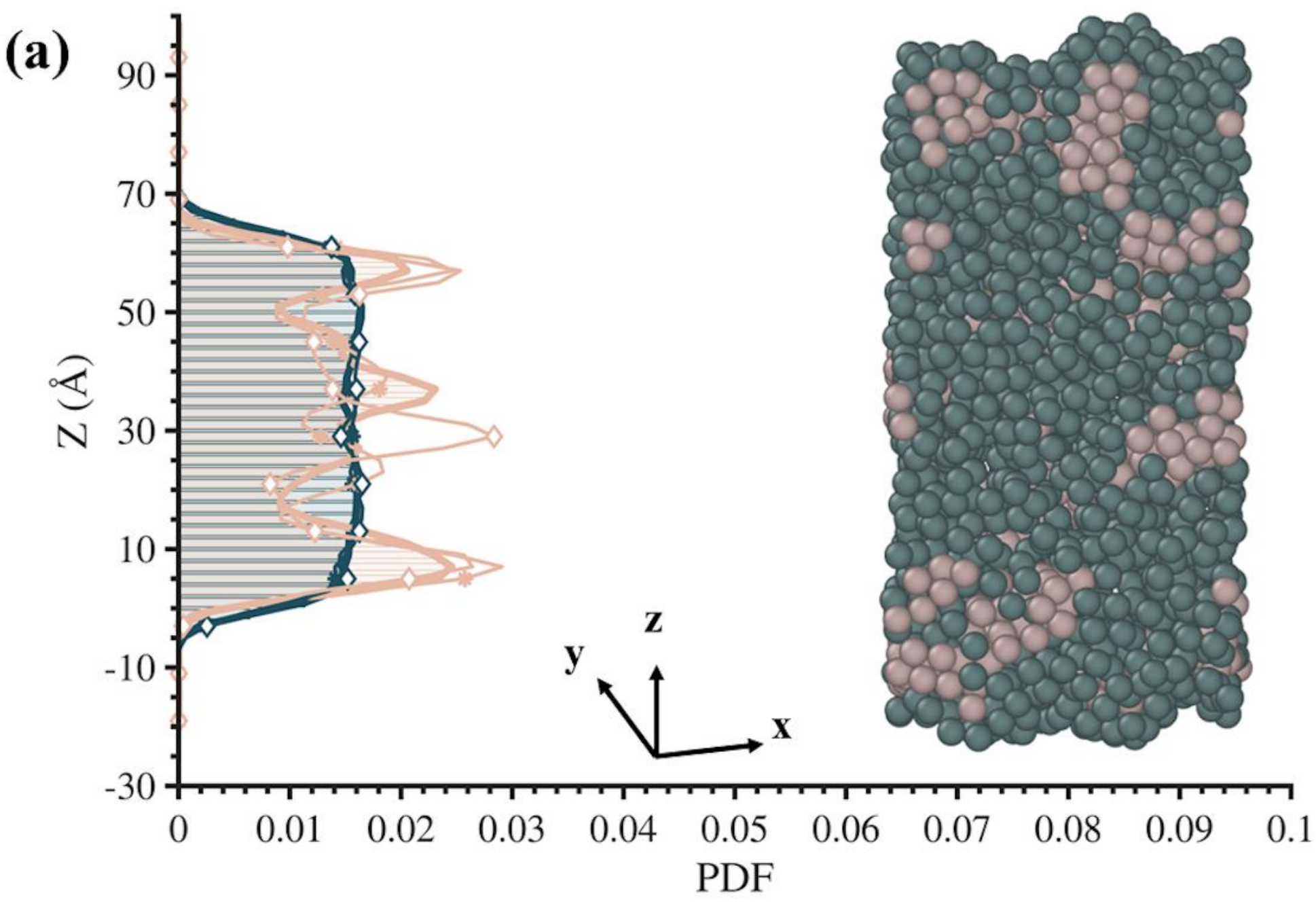
(a)
90
70
50
30
10
-10
-30
Z (Å)
0
0.01
0.02
0.03
0.04
0.05
0.06
0.07
0.08
0.09
0.1
PDF
z
y
x

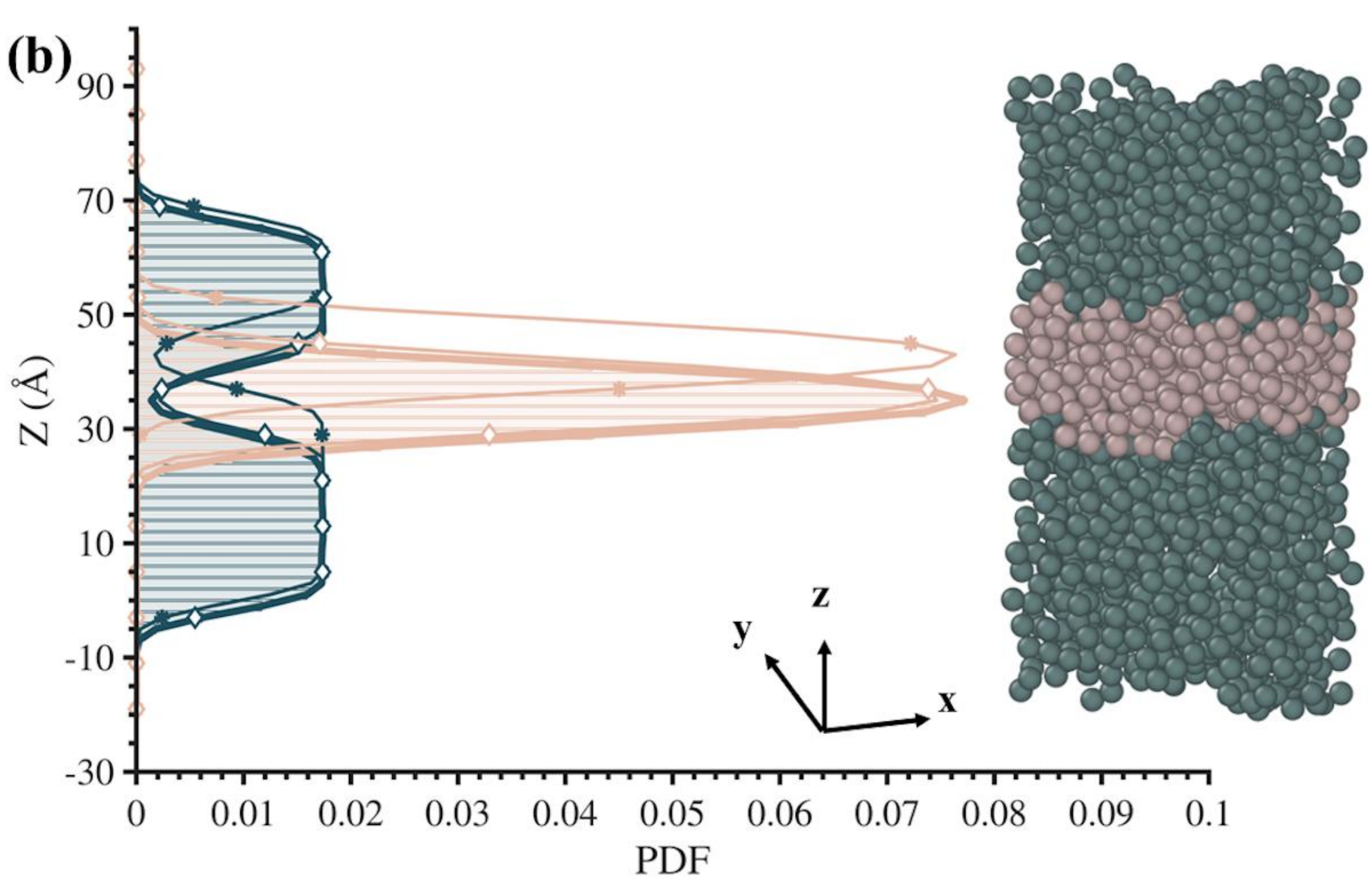
(b)
90
70
50
30
10
-10
-30
Z (Å)
0
0.01
0.02
0.03
0.04
0.05
0.06
0.07
0.08
0.09
0.1
PDF
z
y
x

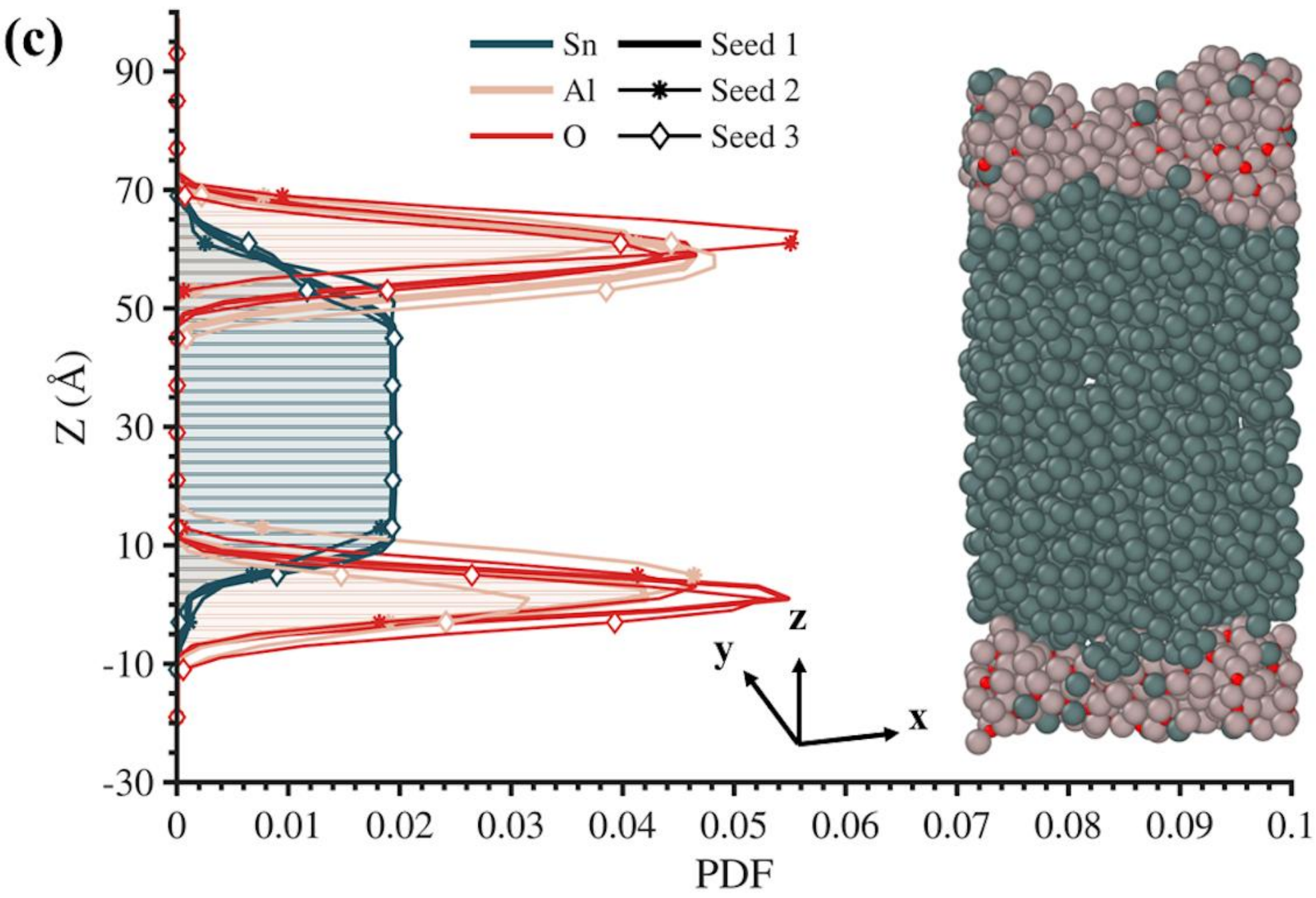

Figure 3. Probability density profiles and atomic configurations of the Sn–Al alloy slab at (a) 300 K and (b) 1300 K, and (c) Sn-Al-O slab at 1300 K. Left panels: Depth-resolved probability density functions (PDFs) of Sn (deep teal), Al (light brown) and O (red) atoms along the *z*-axis, illustrating their spatial distributions across the slab. Right panels: Atomic snapshots showing the mixed Sn-Al and Sn-Al-O structures at each temperature.

Figure 3(b) demonstrates strong chemical separation compared to the more homogeneous distribution at lower temperatures. The probability density functions (PDFs) of Sn and Al, obtained from three independent simulations with different initial velocity seeds (Seeds 1–3), show consistent behavior in which Al forms a sharp, localized peak at the center of the slab, accompanied by a depletion of Sn in the same region. The agreement across seeds confirms the robustness of this segregation behavior. Consistent with classical thermodynamic predictions, this behavior can be rationalized using the Butler equation and Gibbs adsorption theory, where the component with lower surface tension preferentially segregates to interfaces. In the Sn–Al system, Sn has a significantly lower surface tension (approximately half that of Al), driving its enrichment at the free surfaces, in agreement with prior molecular dynamics studies of Sn–Al–Zn liquid alloys.[36] As a result, Sn migrates toward both interfaces in this double-interface system, while Al is displaced toward the center. The atomistic snapshot corroborates these findings, revealing a distinct Al-rich layer confined to the middle of the slab, surrounded by Sn-rich regions on both sides. This clear compositional partitioning reflects segregation driven by atomic diffusion and interfacial energetics and is in excellent agreement with thermodynamic expectations. Overall, the system transitions from a mixed to a strongly segregated state under elevated temperature conditions.

Figure 3(c) combines a one-dimensional PDF with a three-dimensional atomic snapshot to illustrate the spatial distribution of Sn, Al, and O. In the interior region ($z \approx 10$–50 Å), Sn remains dominant, while Al and O exhibit negligible presence, indicating limited solubility within the slab. In contrast, both Al and O display strong enrichment near the free surfaces ($z \approx 0$–10 Å and $z \approx$ 50–65 Å), with overlapping PDF peaks that signify co-segregation and strong chemical coupling. The consistency across different initial velocity seeds further confirms the robustness of this interfacial segregation behavior. The coincident enrichment of Al and O indicates the formation of Al-O rich interfacial layers, consistent with oxygen's stronger affinity for Al compared to Sn, in agreement with prior molecular dynamics studies of the Sn–Al–O system.[37] The corresponding depletion of Sn near oxygen-exposed surfaces demonstrates a segregation-driven redistribution, in which Al effectively acts as an oxygen getter, stabilizing the interface while displacing Sn toward the interior. The atomistic snapshot directly supports this interpretation by revealing oxygen-decorated Al-rich layers localized at both interfaces, while the interior remains structurally uniform and Sn-rich.

Figure 4(a) depicts the atomic percentage profiles of Sn–Al along the slab-normal ($z$) direction at 1300 K, quantifying the spatial distribution of species across the slab. In the surface regions ($z <$ ~20 Å and $z >$ ~50 Å), Sn is strongly enriched, reaching nearly 100 at.%, while Al is almost completely depleted. In contrast, within the interior region ($z \approx 30$–40 Å), Al exhibits a pronounced peak (~90–95 at.%), accompanied by a corresponding minimum in Sn concentration (~5–10 at.%). This symmetric compositional distribution reflects the double-interface nature of the system, with Sn segregating toward both free surfaces and Al confined to the slab interior, consistent with the thermodynamic trends discussed above. The total atom count varies across the slab, with a relatively uniform distribution in the bulk region (~50–60 atoms per bin) and a pronounced increase in the central region (~80–90 atoms per bin), along with noticeable deviations near the interfacial regions. Overall, the results provide clear quantitative evidence of surface-driven segregation and the formation of an Al-rich core.

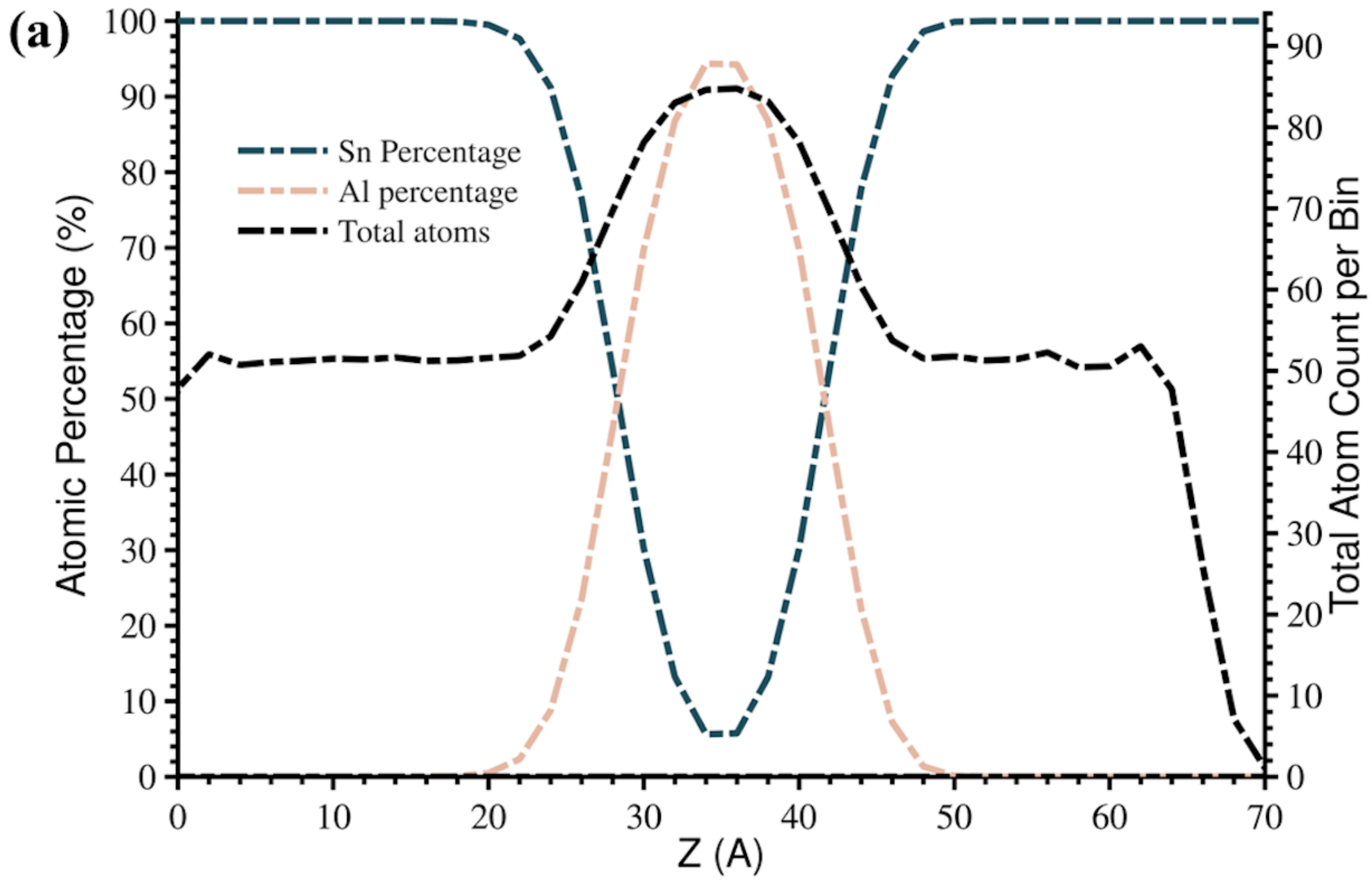

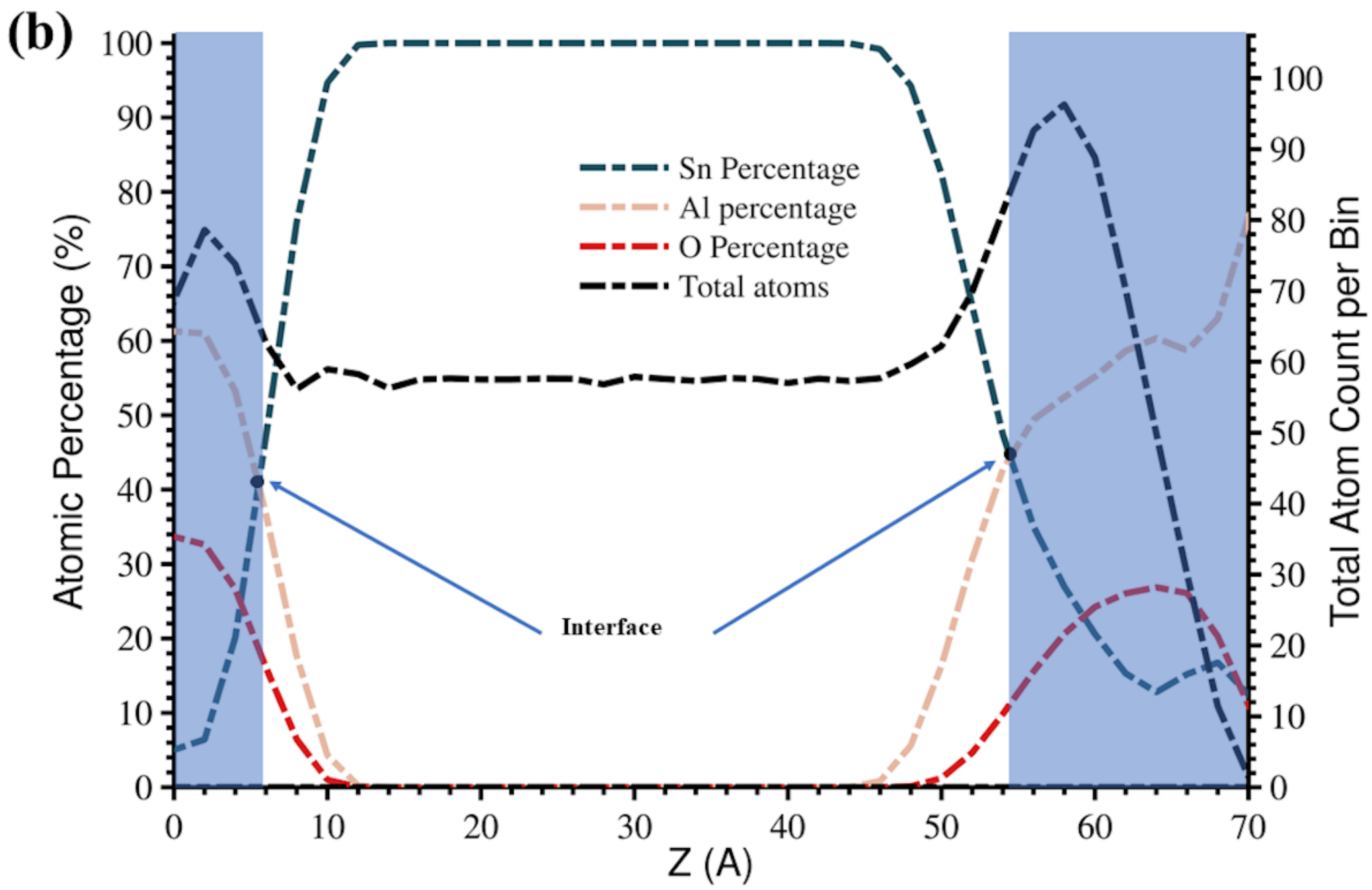

Figure 4. (a) Depth profile of Sn and Al atomic percentages in Sn–Al slab at 1300 K. (b) Same slab after oxygen exposure, showing Sn, Al, and O along the *z*-axis. Oxygen penetrates near-surface regions and creates interfacial zones with abrupt Sn and Al changes, marking reactive boundaries.

The role of oxygen in modifying interfacial chemistry is revealed in Figures 3(c) and 4(b), which extend the analysis to Sn–Al–O systems. Figure 4(b) further quantifies this behavior through atomic percentage profiles. In the interior region ($z \approx 12$–$46$ Å), Sn maintains a nearly constant concentration of ~100 at.%, indicating a well-preserved bulk-like region with negligible penetration of oxygen and aluminum. In contrast, the shaded regions ($z \lesssim 6$ Å and $z \gtrsim 55$ Å), identified here as the interfacial zones, exhibit pronounced compositional gradients over a narrow width of ~5–10 Å. Within these regions, Sn is significantly depleted to ~10–20 at.% at the outermost layers, while Al and O show strong enrichment, reaching peak values of ~60–75 at.% and ~25–30 at.%, respectively. The total atom count also decreases near the boundaries, consistent with the presence of free surfaces. Moreover, the near-symmetric nature of these profiles across both sides of the slab indicates a well-equilibrated double-surface configuration.

Overall, these results demonstrate that while the Sn–Al system can remain relatively uniform in the absence of reactive species, it becomes highly susceptible to segregation under thermodynamic driving forces. Oxygen exposure significantly amplifies this behavior by promoting Al segregation and stabilizing Al–O–rich interfacial layers, while preserving a Sn-rich interior.

## 3.2 Inherent and nonmetal-mediated segregation and interfacial chemistry in liquid Sn–Li alloys

Figure 5 depicts the PDFs comparing the Sn–Li 300 K and 1300 K states, while Figure 6 details the atomic percentage profiles specifically at 1300 K for the Sn–Li slab. In both temperature conditions shown in Figure 5, the Sn and Li PDFs exhibit a substantial overlap across the slab thickness, indicating a largely mixed system without pronounced compositional gradients. At 300 K (Figure 5a), the system remains in an equilibrated solid-like state, and the PDFs display distinct peaks arising from underlying structural ordering. In contrast, at elevated temperature (1300 K, Figure 5b), the system transitions to a liquid state where these structural features smooth out due to enhanced atomic mobility; however, no clear segregation behavior emerges, in stark contrast to the strong compositional separation observed in the Sn–Al system under similar conditions.

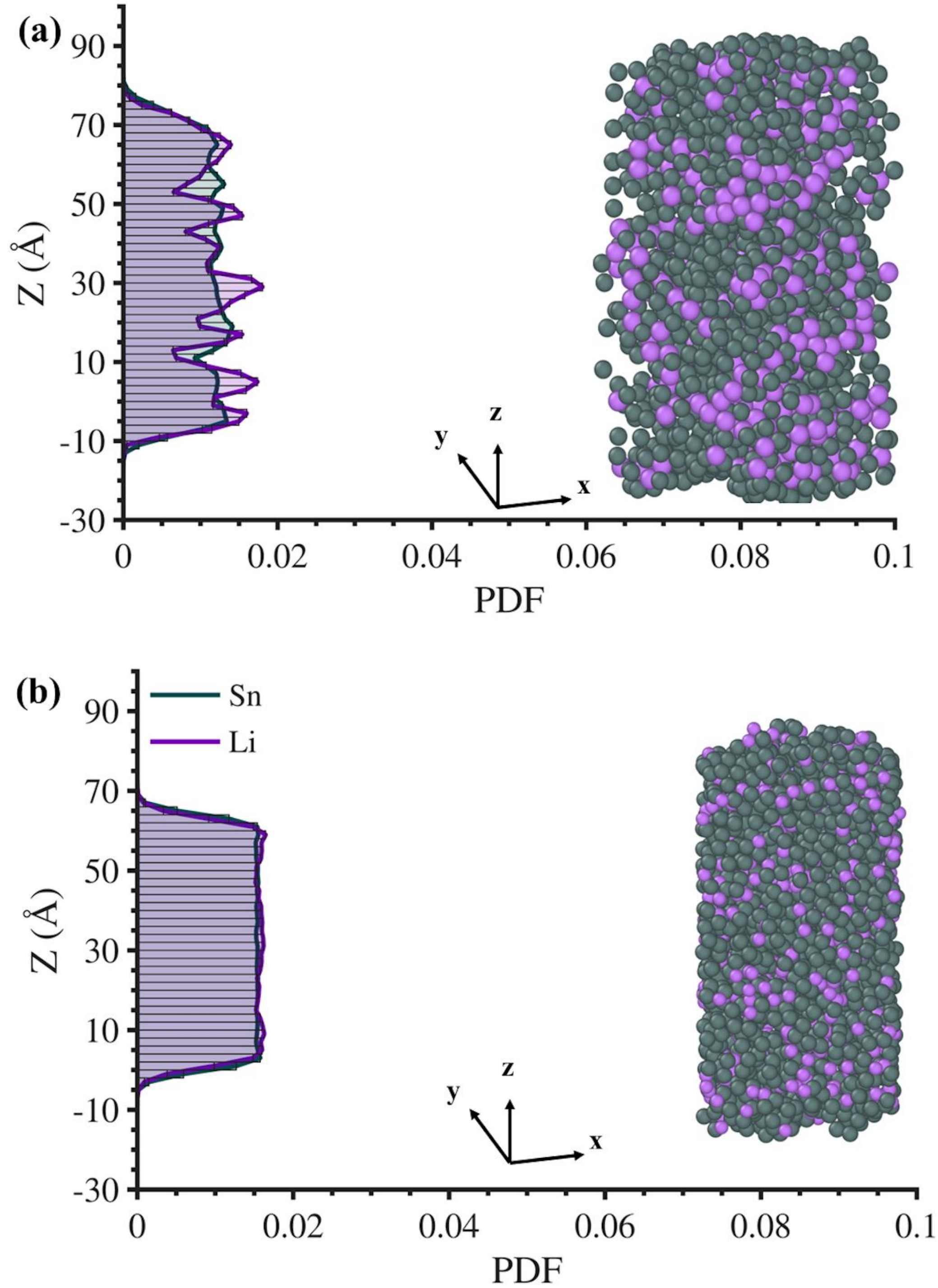

Figure 5. Depth-resolved probability density profiles and atomic configurations of the Sn–Li slab at (a) 300 K and (b) 1300 K. PDFs of Sn (deep teal) and Li (purple) along the *z*-direction are shown (left), with corresponding atomic configurations (right).

The atomic configuration at 1300 K further supports this observation, showing Li atoms uniformly dispersed throughout the Sn matrix without forming clusters or preferential surface enrichment. This homogeneous distribution confirms the absence of a strong thermodynamic driving force for Li segregation. Figure 6 further quantifies this behavior at 1300 K through atomic percentage profiles. Across the slab ($z \approx 5$–60 Å), Sn maintains a nearly constant concentration of ~70–75 at.%, while Li remains at ~25–30 at.%, with only slight variations near the free surfaces ($z \lesssim 5$ Å and $z \gtrsim 60$ Å). The total atom count per bin remains nearly uniform (~55–65 atoms), indicating statistical consistency, and the symmetry of the profiles confirms a well-equilibrated double-interface configuration. These results are consistent with prior *ab initio* molecular dynamics studies by del Rio *et al.*, which also reported no or only mild surface segregation in Sn–Li systems at elevated temperatures.[20] Overall, the Sn–Li alloy exhibits fundamentally weak segregation tendencies, even under conditions that promote atomic diffusion, highlighting a clear contrast with the strongly segregating Sn–Al system.

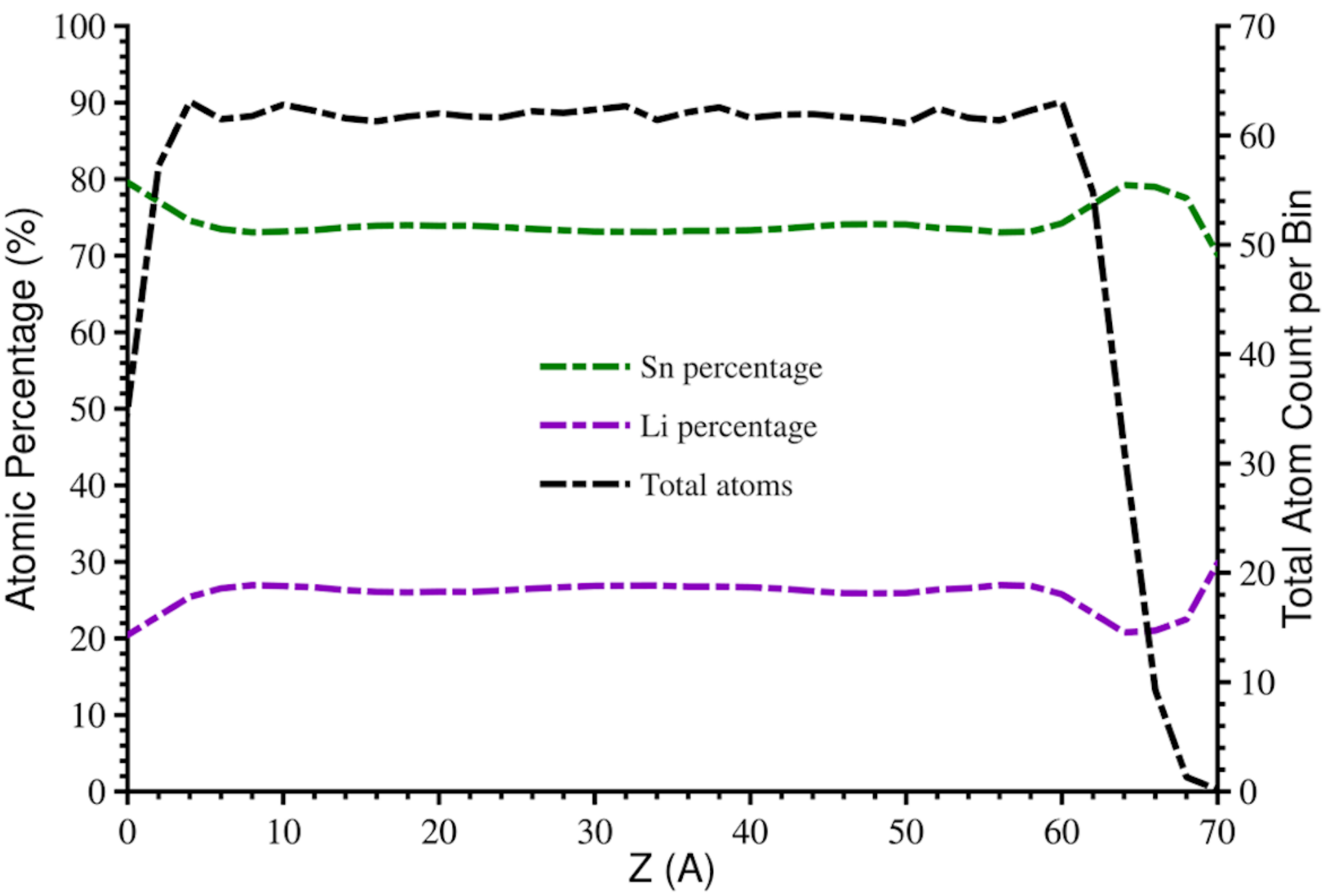

Figure 6. Depth profile of atomic composition in the Sn–Li slab at 1300 K, showing the atomic percentages of Sn and Li as a function of depth along the z-axis.

### 3.2.1 Oxygen-driven interfacial chemistry and segregation

The influence of oxygen on the Sn–Li interfacial chemistry and segregation behavior at 1300 K is depicted in Figures 7 and 8. Figure 7 combines one-dimensional probability density profiles of Sn, Li, and O with a corresponding atomic configuration snapshot, while Figure 8 illustrates the associated atomic percentage profiles. In contrast to the homogeneous Sn–Li distribution observed at 1300 K without oxygen, oxygen exposure induces strong interfacial segregation accompanied by clear Li-rich clustering. In Figure 7, both Li and O exhibit pronounced enrichment near the two free surfaces, centered approximately at $z \approx 0$–20 Å and $z \approx 40$–60 Å, while Sn is depleted in those same regions. The strong overlap between the Li and O PDFs indicates preferential Li–O association. The atomistic snapshot directly supports this interpretation, revealing clustered Li-rich, oxygen-decorated interfacial domains on both sides of the slab, whereas the center remains predominantly Sn-rich.

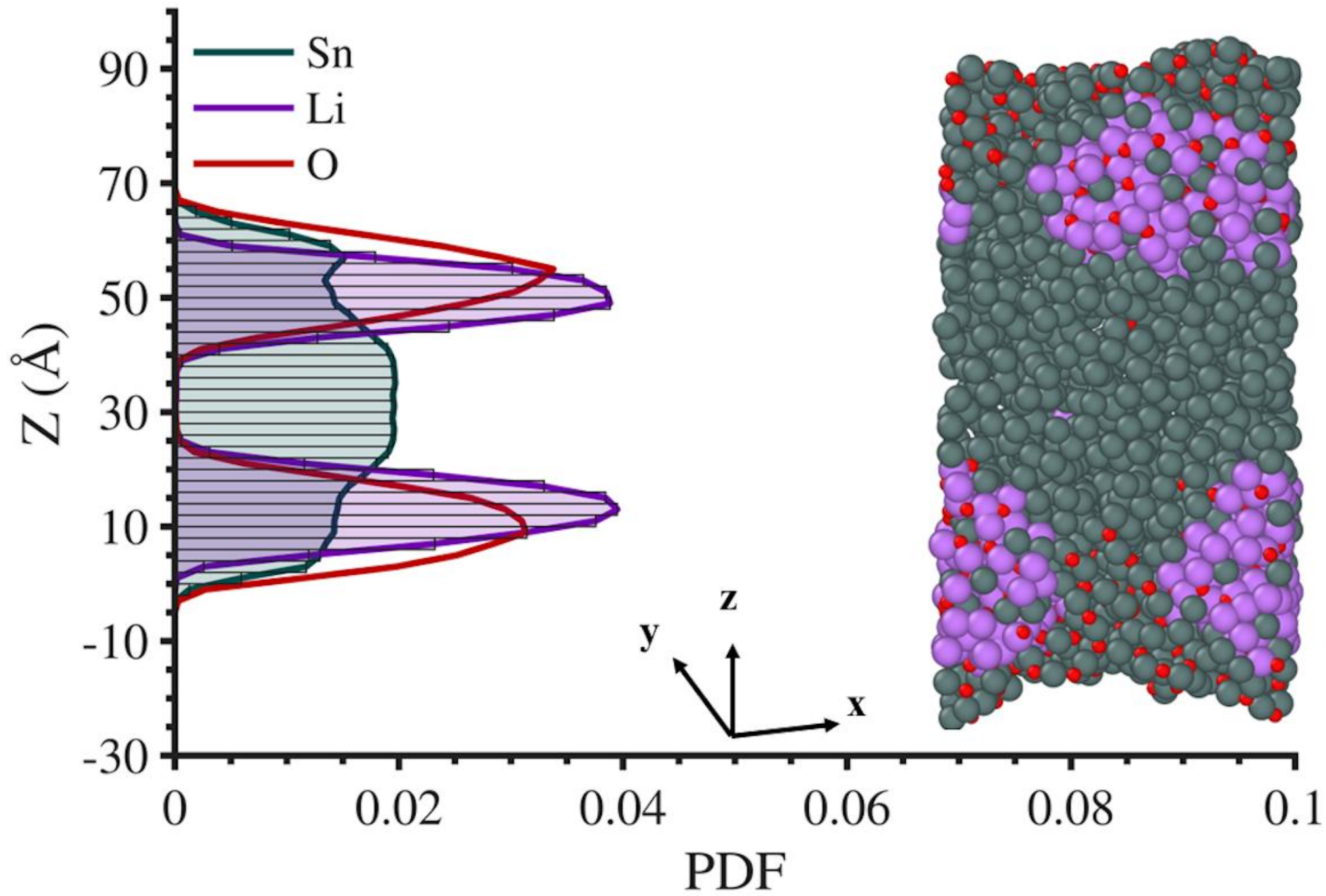

Figure 7. Depth-resolved probability density profiles and atomic configurations of the Sn–Li–O slab at 1300 K. PDFs of Sn (teal), Li (purple), and O (red) along the *z*-direction are shown (left), with corresponding atomic configurations (right).

Figure 8 quantifies this behavior more clearly. The central region of the slab, approximately $z \approx 24$–39 Å, is essentially Sn-rich, where the Sn concentration rises to nearly 100 at.% and both Li and O drop to nearly zero. In contrast, the interfacial Li-rich regions are localized around $z \approx 8$–20 Å and $z \approx 44$–58 Å, where Li reaches roughly 30–35 at.% and oxygen remains strongly enriched. Within these same regions, Sn decreases substantially to approximately 35–40 at.%, confirming strong interfacial chemical partitioning. The total atom count per bin is lower and nearly flat in the Sn-rich center, but higher in the Li-rich interfacial zones, consistent with local clustering and restructuring near the surfaces.

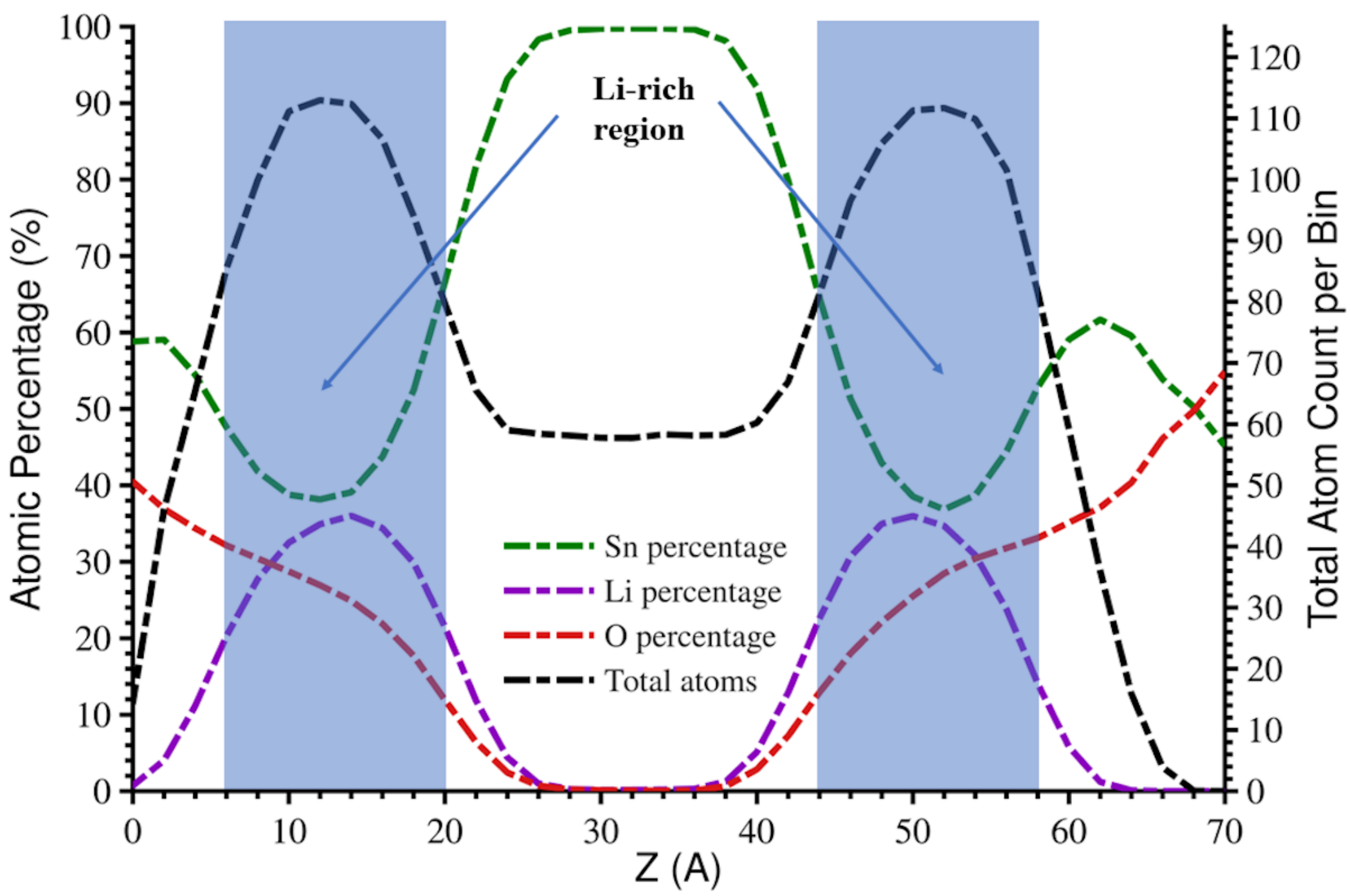

Figure 8. Depth-resolved atomic composition of the oxygen-exposed Sn–Li slab at 1300 K. Atomic percentages of Sn (teal), Li (purple), and O (red) are plotted along the *z*-direction, with the total atom count per bin shown on the secondary axis (black). Shaded regions indicate Li-rich zones, while the central region corresponds to the Sn-rich bulk.

Compared with the oxygen-free Sn–Li system at 1300 K, where Li remains broadly dispersed, and no segregation is observed, oxygen clearly drives Li toward the interfaces through strong Li–O interactions. This behavior also differs from the Sn–Al–O system at 1300 K, where oxygen stabilizes Al-rich surface layers while maintaining a nearly pure Sn bulk. In the present Sn–Li–O case, oxygen promotes the formation of clustered Li-rich interfacial layers, while the slab center becomes almost completely Sn-rich. Overall, these results show that oxygen fundamentally alters segregation behavior in Sn–Li by stabilizing Li-rich interfacial domains and producing a strongly Sn-rich central region.

### 3.2.2 Role of hydrogen in interfacial chemistry and segregation

The influence of hydrogen on Sn–Li interfacial chemistry at 1300 K is illustrated in Figures 9 and 10, where Figure 9 combines one-dimensional probability density functions (PDFs) with an atomistic snapshot, and Figure 10 presents the corresponding atomic percentage profiles. The presence of hydrogen induces a distinct but moderate segregation pattern characterized by interfacial enrichment of both Li and H. In Figure 9, the PDFs exhibit pronounced peaks for Li and H near the two free surfaces, located approximately at $z \approx 0$–12 Å and $z \approx 50$–65 Å. In these regions, Sn is correspondingly depleted, while remaining dominant in the slab interior. The spatial overlap between Li and H distributions indicates a strong Li–H affinity, consistent with several prior studies that have reported the formation and stability of Li–H bonds.[38–41] The atomistic snapshot supports this interpretation, showing Li-enriched interfacial regions decorated with hydrogen atoms, forming loosely clustered Li–H-associated domains near both surfaces, while the interior remains relatively homogeneous and Sn-rich.

Figure 10 quantitatively captures this behavior. In the interior region ($z \approx 20$–50 Å), Sn maintains a high concentration of ~80–85 at.%, while Li remains at ~20 at.%, confirming a predominantly Sn-rich liquid matrix. Hydrogen concentration in this region is nearly zero, indicating strong interfacial localization. In contrast, the interfacial regions ($z \lesssim 10$ Å and $z \gtrsim 55$ Å) exhibit clear compositional gradients, where Sn decreases significantly, while Li increases to ~35–40 at.% and hydrogen rises to ~20–30 at.%. These gradients are confined within narrow regions (~2–8 Å), indicating localized segregation. The total atom count per bin remains nearly uniform in the interior, and the symmetry about the slab center confirms a well-equilibrated double-interface configuration.

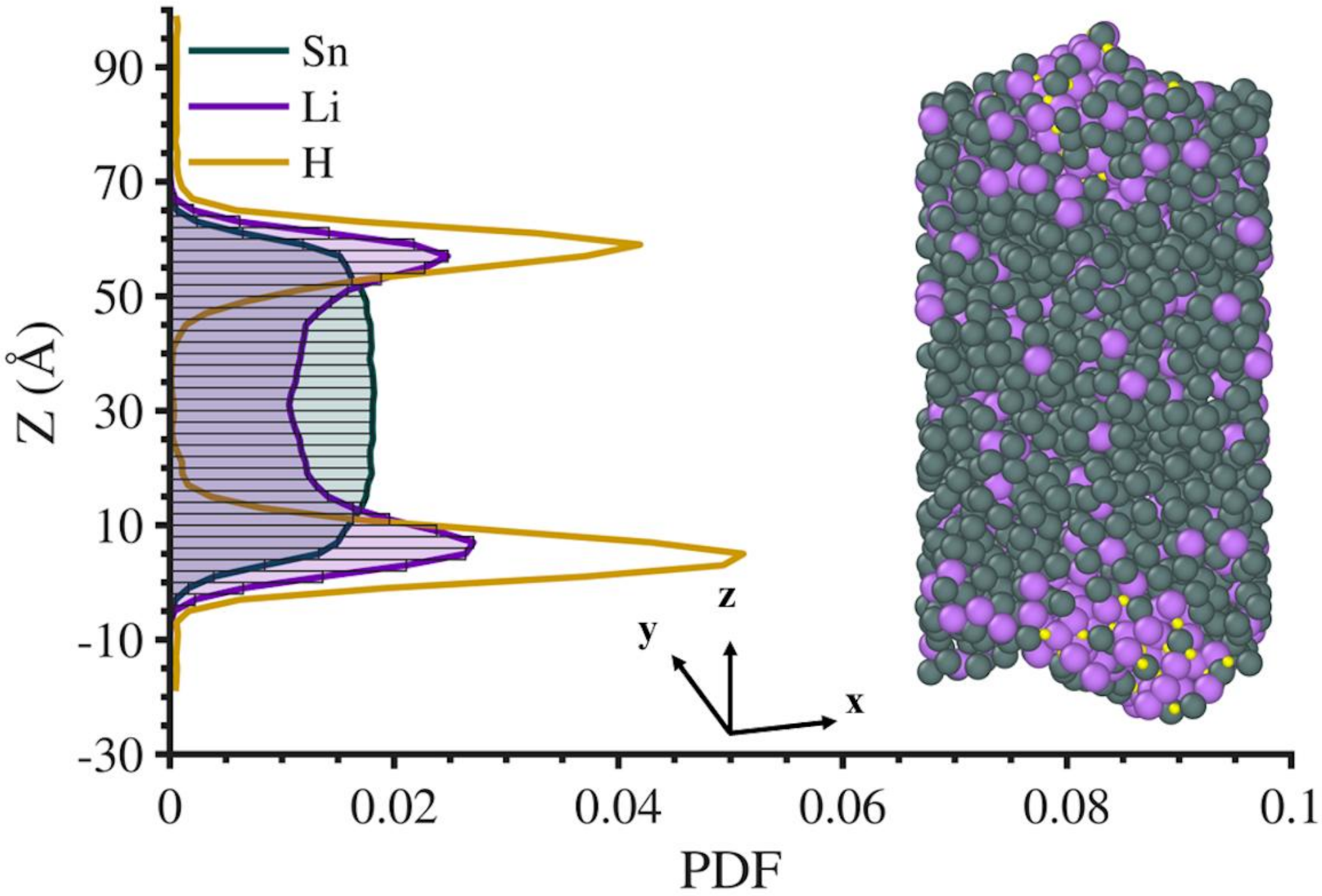

Figure 9. Probability density profiles and atomic configurations of the Sn-Li-H alloy slab at 1300 K. Left panels: Depth-resolved probability density functions (PDFs) of Sn (teal), Li (Purple), and H (yellow) atoms along the *z*-axis, illustrating their spatial distributions within the slab. Right panels: Atomic snapshots depicting the mixed Sn-Li-H structures at 1300 K.

Compared to the oxygen-driven Sn–Li–O system, where strong Li–O interactions produce highly pronounced segregation and well-defined Li-rich clusters, hydrogen induces a comparatively weaker but still distinct segregation response. Unlike the oxygen case, the bulk region in Sn–Li–H remains more mixed and retains a significant Li presence rather than forming a nearly pure Sn core. Overall, these results demonstrate that hydrogen modifies the interfacial chemistry of Sn–Li alloys by stabilizing Li at the surfaces through Li–H interactions, leading to the formation of moderately enriched Li–H interfacial layers while preserving a Sn-rich interior.

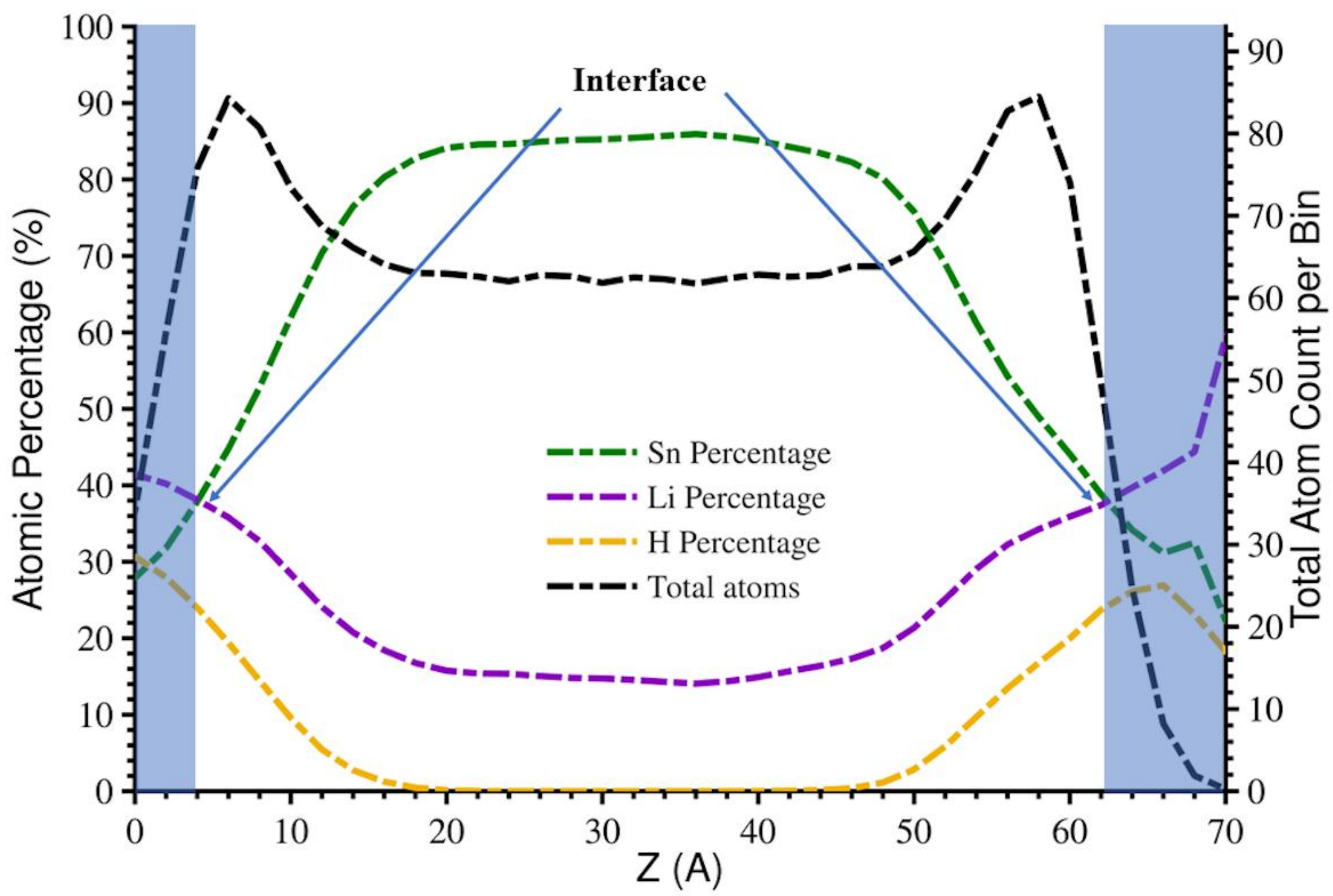

Figure 10. Depth-resolved atomic composition of the hydrogen-exposed Sn–Li slab at 1300 K. The atomic percentages of Sn, Li, and H are plotted along the *z*-axis, highlighting hydrogen penetration into near-surface regions (shaded areas) and the formation of interfacial zones where Sn and Li concentrations change abruptly. These interfacial regions represent reactive boundaries that form upon hydrogen exposure.

### 3.2.3 Coupled oxygen-hydrogen effects on interfacial chemistry and surface segregation

The combined influence of hydrogen and oxygen on Sn–Li interfacial chemistry at 1300 K is illustrated in Figures 11 and 12, where Figure 11 presents the PDFs and an atomistic snapshot, and Figure 12 shows the corresponding atomic percentage profiles. Compared to the oxygen-only and hydrogen-only systems, the simultaneous presence of both species induces a significantly enhanced and more localized segregation response. In Figure 11, the PDFs show strong enrichment of Li, O, and H near both free surfaces, centered approximately at $z \approx$ -10–10 Å and $z \approx$ 50–75 Å, while Sn is strongly depleted in these regions. A pronounced overlap among Li, O, and H distributions indicates cooperative interactions, consistent with the formation of Li–O–H-associated complexes at the interfaces. The atomistic snapshot clearly supports this interpretation, revealing densely packed, clustered interfacial domains enriched with Li, O, and H, forming chemically complex surface layers on both sides of the slab. In contrast, the interior region remains relatively uniform and dominated by Sn.

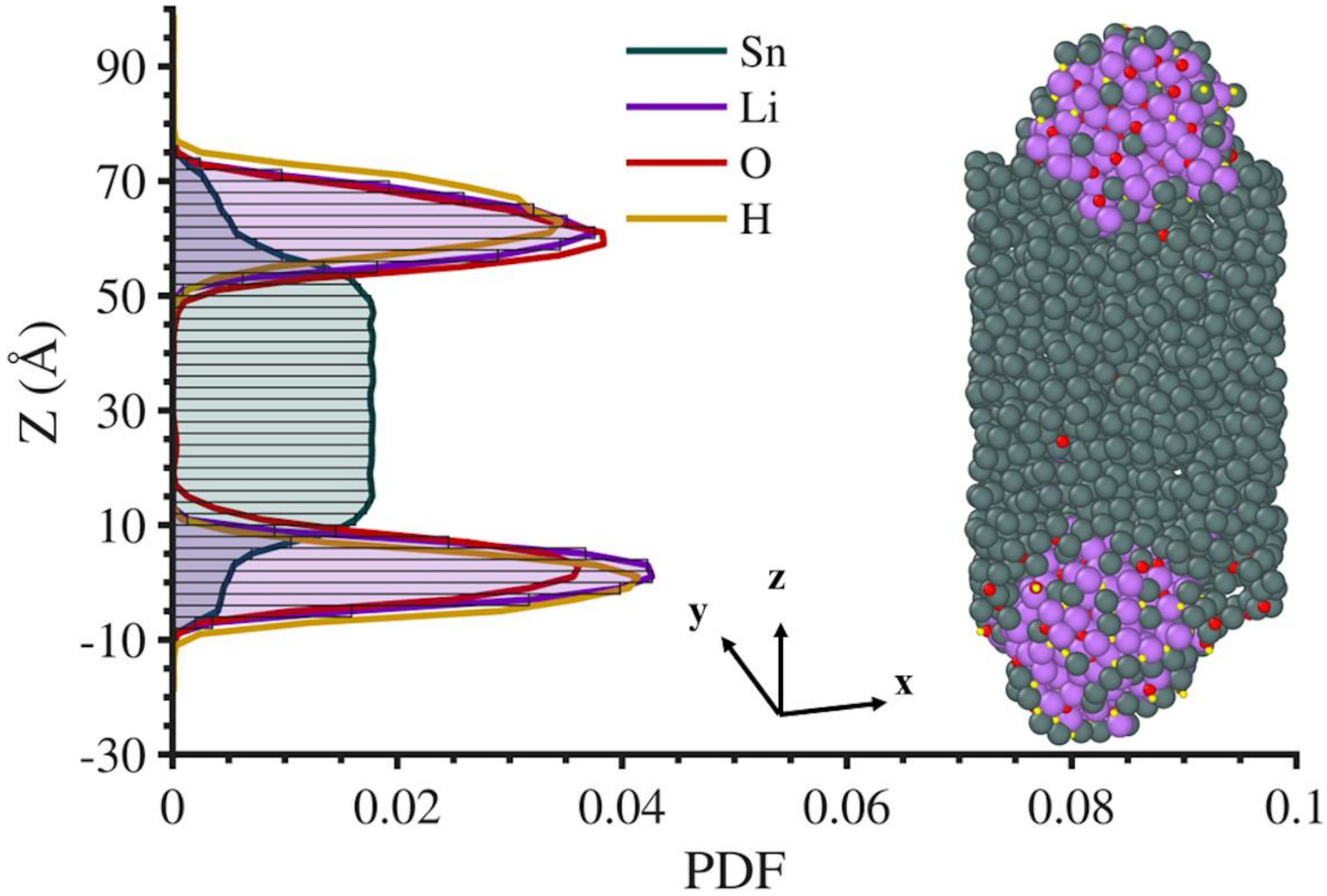

Figure 11. Depth-resolved probability density profiles and atomic snapshots of the Sn–Li–O–H slab at 1300 K. PDFs of Sn (teal), Li (purple), O (red), and H (yellow) along the *z*-direction are shown (left), with corresponding atomic snapshots (right).

Figure 12 quantitatively confirms this behavior. The central region of the slab ($z \approx 12$–50 Å) becomes nearly pure Sn, where the Sn concentration approaches ~95–100 at.%, while Li is reduced to only trace amounts (~0–5 at.%) and both O and H are nearly absent. This indicates strong expulsion of reactive species from the bulk due to oxygen and hydrogen-driven surface segregation. In contrast, the interfacial regions ($z \lesssim 10$ Å and $z \gtrsim 55$ Å) exhibit sharp compositional gradients over a narrow width (~8–12 Å), where Sn decreases significantly and Li, O, and H increase simultaneously. Near the interfaces, Li reaches ~35–45 at.%, oxygen ~15–20 at.%, and hydrogen ~15–25 at.%, confirming a synergistic segregation mechanism driven by combined Li–O and Li–H interactions. The total atom count per bin remains consistent in the slab interior, and the symmetric profiles confirm a well-equilibrated double-interface configuration.

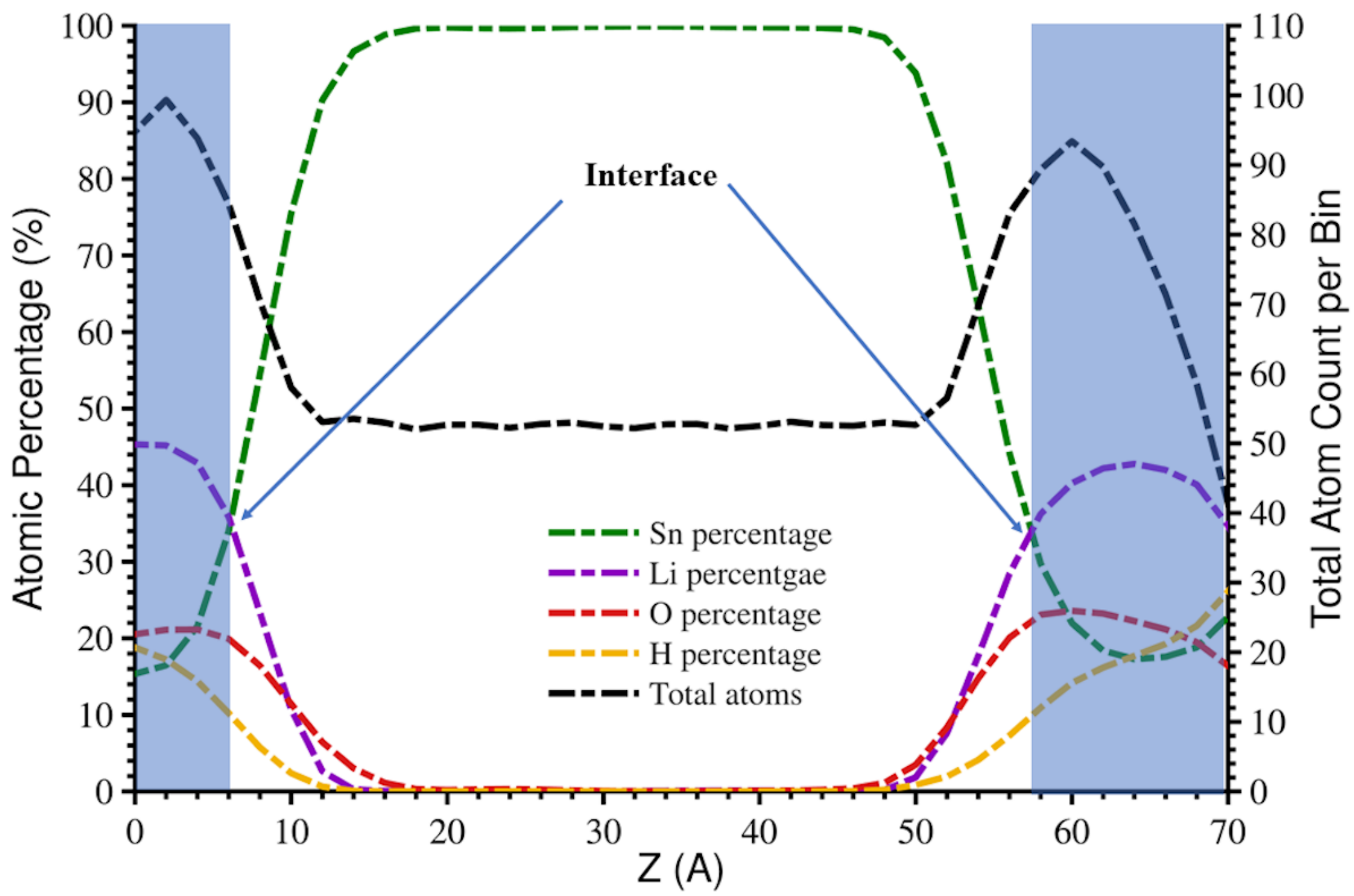

Figure 12. Depth-resolved atomic composition of the combined oxygen- and hydrogen-exposed Sn–Li slab at 1300 K. The atomic percentages of Sn, Li, O and H are plotted along the *z*-axis, highlighting oxygen and hydrogen penetration into near-surface regions (shaded areas) and the formation of interfacial zones where Sn and Li concentrations change abruptly. These interfacial regions represent reactive boundaries that form upon oxygen and hydrogen exposure.

Compared to the Sn–Li–H system, where the bulk remains partially mixed, and the Sn–Li–O system, where Li segregation is strong while Sn remains dominant in the bulk, the combined Sn–Li–O–H system exhibits the most pronounced surface segregation behavior, producing a nearly pure Sn core with highly enriched multi-component surface layers. This highlights the dominant role of oxygen in driving segregation, with hydrogen further enhancing interfacial stabilization through additional chemical interactions. Notably, similar segregation trends are also observed for the 9 at.% Li composition across all the studied systems, indicating that the underlying segregation mechanisms remain robust even at lower Li concentrations. Overall, these results demonstrate that the combined presence of oxygen and hydrogen dramatically alters the interfacial chemistry of liquid Sn–Li alloys by promoting the formation of clustered Li–O–H-rich interfacial layers while expelling reactive species from the bulk.

### 3.2.4 Benchmarking

To validate the physical relevance of the observed surface segregation behavior, the simulation results were compared with experimental measurements by Bastasz and Whaley,[23] particularly their Figure 10, which shows that lithium preferentially segregates to the surface of Sn–Li alloys and reaches ~50% surface coverage at elevated temperatures (≈450–500 °C) in the presence of oxygen. In the present work, simulations were performed for the same alloy composition (Sn–Li, 20 at% Li) at 800 K (≈527 °C), 1000 K, and 1300 K, enabling a direct comparison within and beyond the experimental temperature range. At 800 K, which closely corresponds to the experimental conditions, both free surfaces exhibit Li-rich layers with surface coverage approaching ~45% Li (Figure 13), in good agreement with the reported measurements, noting that the experimental observations correspond to a single free surface. This Li-enriched surface behavior persists at higher temperatures, indicating robust segregation across a broader thermal range.

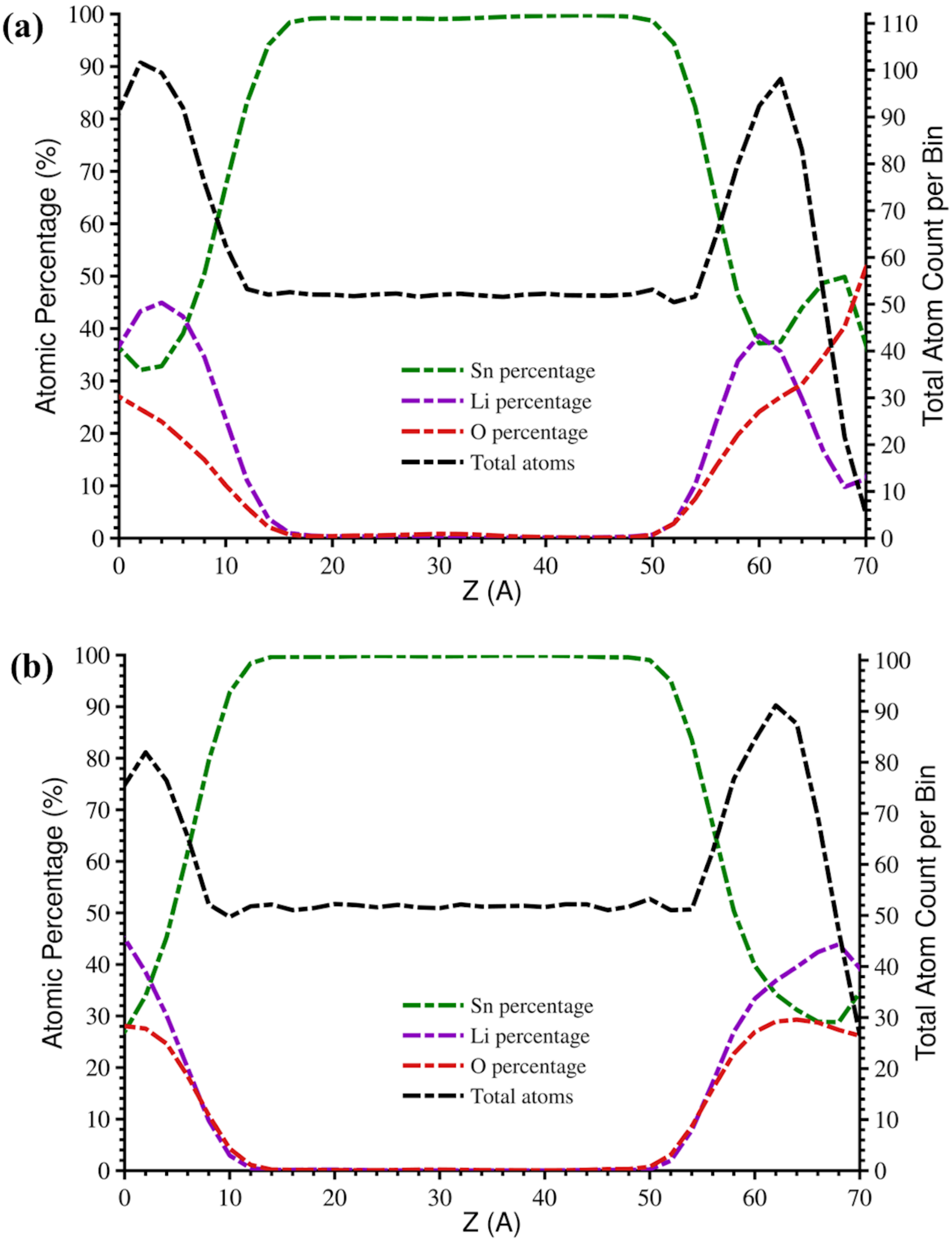
(a)
Atomic Percentage (%)
Total Atom Count per Bin
Z (A)
Sn percentage
Li percentage
O percentage
Total atoms
(b)
Atomic Percentage (%)
Total Atom Count per Bin
Z (A)
Sn percentage
Li percentage
O percentage
Total atoms

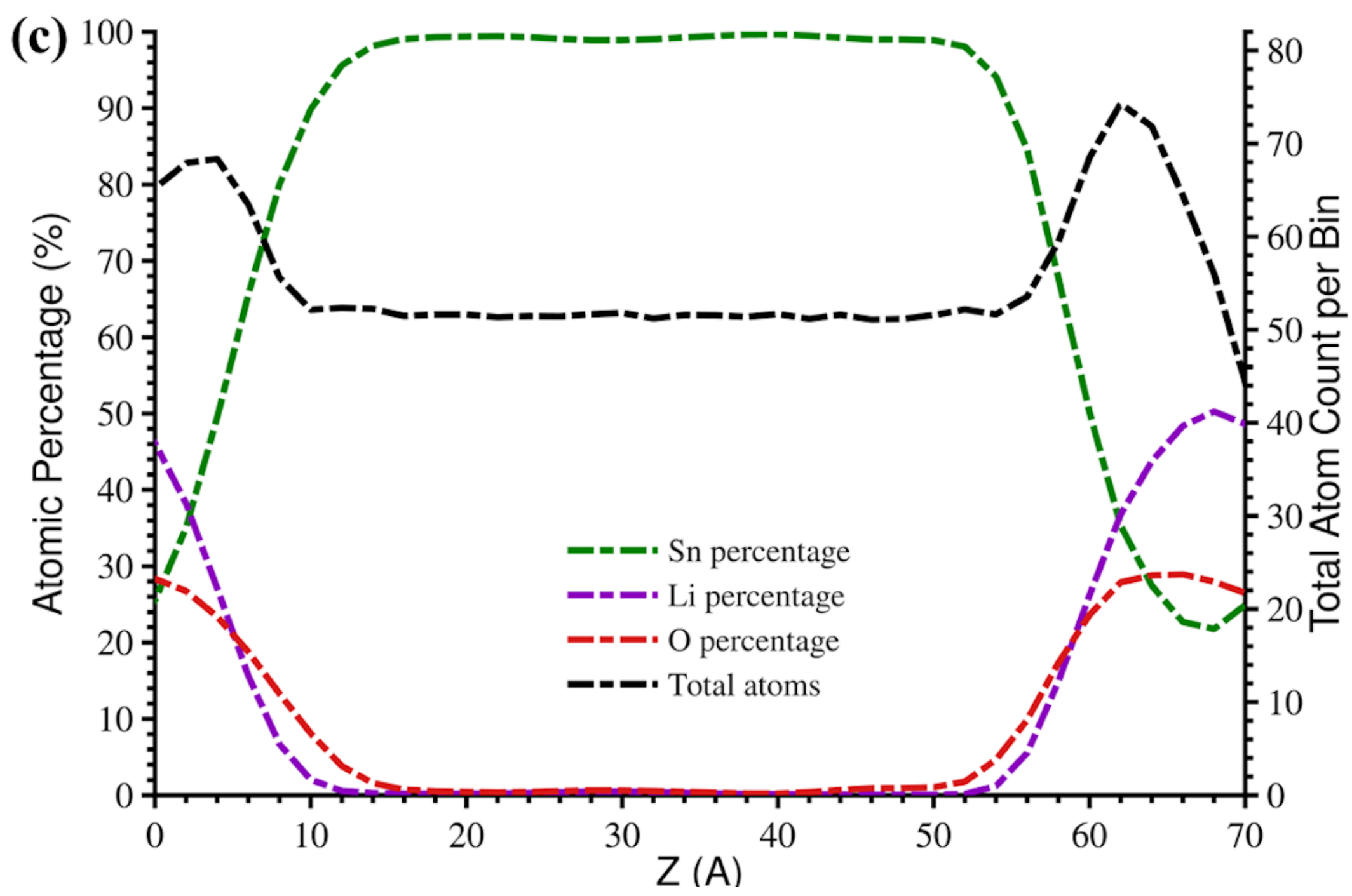

Figure 13. Atomic percentage profiles along the surface-normal direction (*z*-axis) for the Sn–Li–O-H system at (a) 800 K, (b) 1000 K, and (c) 1300 K, showing the distributions of Sn, Li, and O.

Although both oxygen and hydrogen were included in the simulations to represent fusion-relevant environments, the experimental comparison is based on oxygen-containing profiles to maintain consistency with LEIS measurements. Notably, the simulations further show that the combined presence of oxygen and hydrogen enhances segregation relative to oxygen-only conditions, leading to more pronounced Li enrichment at the interfaces. Overall, the strong agreement with experimental trends supports the validity of our models and confirms that the observed segregation behavior is governed by Gibbsian thermodynamics, wherein species with lower surface tension preferentially occupy interfacial regions, while extending these observations to higher, fusion-relevant temperatures beyond the experimental regime.

### 3.3Segregation strength quantification of different plasma-facing alloys

To quantitatively assess interfacial segregation behavior across all investigated systems, a global segregation metric was employed based on the spatial overlap of species distributions along the surface-normal direction. Using Eq. (1), the overlap parameter $S$ was evaluated using the time-averaged atomic density profiles of the two primary alloy constituents, and the corresponding segregation strength was then defined as $1 - S$. In this formulation, a value of $S = 1$ corresponds to complete spatial overlap, representing a fully mixed system, whereas lower values indicate increasing degrees of compositional separation.

The computed segregation strengths for all systems are presented in Figure 14 and categorized in Table 2. Based on this classification, the systems can be divided into distinct segregation regimes. The Sn–Li system at 1300 K exhibits a segregation index approaching zero, indicating a fully mixed state with negligible interfacial enrichment. Similarly, both Sn–Al (300 K) and Sn–Li (300 K) fall within the near-fully-mixed regime ($1 - S < 0.1$), reflecting homogeneous atomic distributions at low temperature in the absence of reactive species. A transition to weak segregation is observed in the Sn–Li–H system at 1300 K, where the segregation index lies in the range of 0.1 – 0.3. This indicates that hydrogen induces mild compositional perturbations, leading to slight interfacial enrichment without significant phase separation. In contrast, the Sn–Li–O system exhibits moderate segregation ($1 - S = 0.3 - 0.7$), demonstrating that oxygen has a stronger influence on interfacial chemistry, promoting noticeable compositional gradients and partial segregation.

Table 2. Classification of segregation strength based on the global segregation index ($1 - S$) and corresponding Sn–Al and Sn–Li alloy systems.

| Segregation Index (1 − S) | Segregation Regime | Alloys in This Range | Physical Interpretation |
|---|---|---|---|
| 0 | Fully Mixed | Sn–Li (1300 K) | Complete overlap of species; no segregation |
| < 0.1 | Near-Fully-Mixed | Sn–Al (300 K), Sn–Li (300 K) | Nearly homogeneous distribution |
| 0.1 – 0.3 | Weak Segregation | Sn–Li–H (1300 K) | Slight enrichment at interfaces |

| Segregation Index (1 − S) | Segregation Regime | Alloys in This Range | Physical Interpretation |
|---|---|---|---|
| 0.3 – 0.7 | Moderate Segregation | Sn–Li–O (1300 K) | Noticeable compositional separation |
| 0.7 – 1.0 | Strong Segregation | Sn–Al (1300 K), Sn–Al–O (1300 K), Sn–Li–O–H (1300 K) | Strong interfacial enrichment / phase separation |

The most pronounced segregation behavior is observed in Sn–Al (1300 K), Sn–Al–O (1300 K), and Sn–Li–O–H (1300 K) alloys, all of which lie within the strong segregation regime ($1 - S > 0.7$). These results indicate significant interfacial enrichment in Sn–Al, the formation of chemically distinct surface layers in Sn–Al–O, and strong compositional separation in Sn–Li–O–H. The enhanced segregation in the Sn–Li–O–H system suggests a cooperative effect between oxygen and hydrogen, leading to complex interfacial structures and amplified segregation compared to single-species environments.

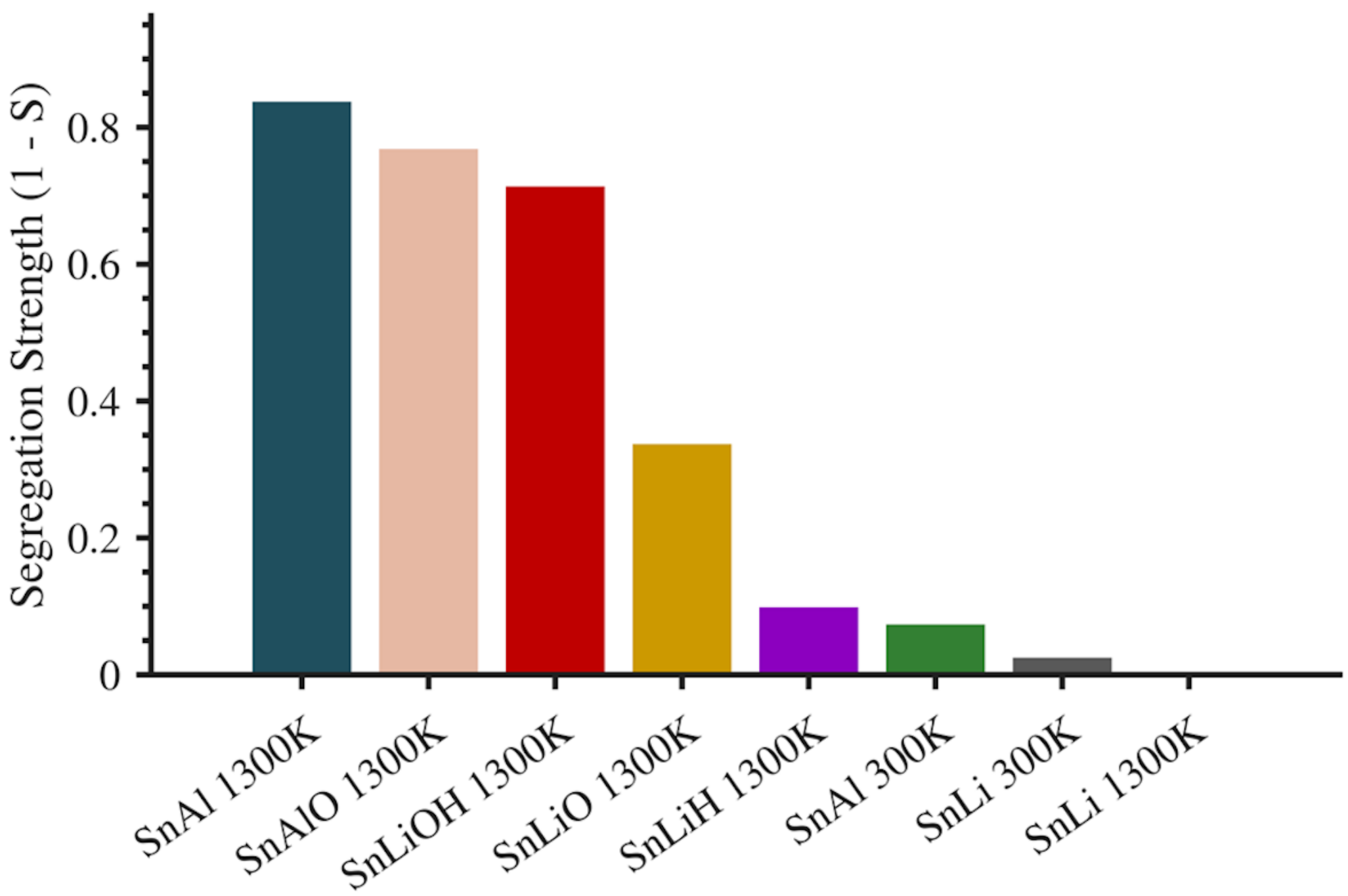

Figure 14. Global segregation strength, defined as ($1 - S$) (where $S$ is defined in Eq. (1)), for all investigated Sn–Al and Sn–Li alloy systems under different chemical environments and temperatures. The systems are ranked in descending order of segregation strength.

Overall, the results presented in the preceding sub-Sections demonstrate that temperature and the presence of non-metallic species, particularly oxygen, play a critical role in governing strong segregation behavior in Sn–Al alloys. While baseline Sn–Li systems remain largely homogeneous, the introduction of oxygen and hydrogen significantly enhances segregation, with combined O–H environments producing the most pronounced surface segregation. This trend is consistently captured by the overlap-based segregation index, confirming its robustness as a quantitative descriptor of interfacial segregation in multi-component alloy systems.

## 4. Conclusions

In this investigation, we used atomistic ReaxFF simulations to model surface segregation in liquid Sn-based plasma-facing alloys, with particular focus on the effects of alloy composition and reactive non-metal species (O and H) on interfacial chemistry. We developed and validated a Sn/Al/Li/O/H ReaxFF force field against DFT-based formation energies, convex-hull thermodynamics, and elastic constants, showing its suitability for large-scale segregation studies. For the binary alloys, we observed that Sn–Al shows strong thermodynamically driven segregation at high temperature: Sn enriches the free surfaces, while Al is displaced inward, consistent with Gibbs adsorption and Butler-equation expectations. In contrast, Sn–Li remained largely mixed even at elevated temperature, indicating that Li has only a weak intrinsic tendency to segregate in the absence of reactive impurities. A notable observation was that oxygen and hydrogen dramatically changed segregation behavior. Oxygen strongly promotes segregation by stabilizing the more reactive alloying element at the interface: in Sn–Al–O, oxygen drives Al-rich surface layers while preserving a Sn-rich bulk, whereas in Sn–Li–O, oxygen pulls Li toward the surfaces, producing Li-rich interfacial regions and a Sn-rich central region. Hydrogen also promotes segregation, but weaker, mainly by stabilizing Li near the interfaces through Li–H interactions.

When both oxygen and hydrogen are present, the segregation effect becomes strongest in the Sn–Li–O–H system. The combined chemistry produces Li–O–H-rich surface layers and an almost pure Sn core, indicating that oxygen is the dominant driver while hydrogen further enhances interfacial stabilization. To quantify these effects, we introduced a global overlap-based segregation metric, which weighs how strongly species separate along the surface-normal direction. This metric provided a unified way to compare segregation across systems and confirms a clear hierarchy.

## 5. Acknowledgments

This work was supported by the U.S. Department of Energy (Award Number: DE-AR0001995) in collaboration with ExoFusion Inc.

# Supporting information

# Surface segregation of liquid metal plasma-facing component alloys: A ReaxFF investigation

Md Adnan Mahathir Munshi,[1] Abdul Aziz Shuvo,[1] Mike Kotschenreuther,[2] Adri C.T. van Duin,[1] and Bladimir Ramos-Alvarado[1*]

[1]Department of Mechanical Engineering, The Pennsylvania State University, University Park, PA-16802, USA.

[2]ExoFusion, Bellevue, WA-98005, USA.

*Corresponding author: E-mail: bzr52@psu.edu

## 1. ReaxFF force field method and Sn/Al/Li/H/O parameterization

Section 1(a) describes the development of an initial ReaxFF description for the Al/Sn/O/H chemistry, whose training set was organized into four components: Al metal, Sn and Sn–O, Al-oxides, and the Al/Sn alloy, and subsequently extended to the Sn-Al-O ternary oxide chemistry. Section 1(b) describes the further extension of the force field to Sn/Li/O/H, with the previously validated Sn/Al/O/H parameters held fixed or allowed only narrow variation, while the Li-containing atomic, bond, off-diagonal, valence-angle, and torsion parameters were released for optimization. This hierarchical release, consistent with the framework of van Duin *et al.*,[1] preserves the transferability of the parameters fixed in the earlier step and avoids simultaneous fitting of strongly correlated parameter subsets.

## 1(a) Development of the Sn/Al/O/H ReaxFF force field description:

The initial training set for the Al/Sn/O/H chemistry was organized into four complementary components: Al metal, Sn and Sn-O, Al-oxides (including hydroxides and aluminum hydrides), and the Al/Sn alloy, and subsequently extended with Sn-Al-O ternary oxide phases. The individual components are described below. Five Al polymorphs (fcc, bcc, sc, A15, and diamond-cubic) and seven Sn polymorphs (α-Sn, β-Sn, bcc, fcc, hcp, and $Sn_8$ simple cubic) were included at their optimized geometries with a comprehensive equation-of-state (EOS) series to anchor the metallic bonding across diverse coordination environments. For Al, explicit training targets included the fcc lattice parameter (ReaxFF 4.037 Å vs. 4.016 Å reference), relative polymorph energies, cohesive energy (ReaxFF −79.08 vs. −78.10 kcal $mol^{-1}$ reference), and vacancy formation energy (ReaxFF 20.72 vs. 15.62 kcal $mol^{-1}$ reference) to constrain both bulk properties and point-defect energetics.

Building on these elemental references, the oxide chemistry was parameterized through multiple phases. For Sn–O bonding, three $SnO_2$ polymorphs (tetragonal/rutile, Imma, Pnnm) and two SnO polymorphs (tetragonal, Cmce) were included alongside bond-dissociation curves, angle scans, and strain scans. For Al–O bonding, six $Al_2O_3$ polymorphs (corundum, $Rh_2O_3$-type, κ-phase, Cmcm, perovskite, and a low-density variant) were incorporated with EOS scans, supplemented by molecular clusters ($Al_4O$, $Al_4O_4$, $Al_4O_6$, $Al_6O_{12}$, $Al_8O_{12}$) to sample coordination environments outside bulk crystalline phases. To extend the Al–O description to protonated environments, aluminum hydroxides ($Al_2O_4H_4$, $Al_2O_5H_4$, $Al_2O_6H_6$, $Al(OH)_3$, $Al(OH)_3 \cdot H_2O$), hydrides ($AlH_2$, $AlH_3$, $Al_4H_6$, $Al_3H_5O$, $Al_2H_4O_2$), and $Al(OH)_n$ coordination references (n = 3–6) were included with extensive angle and bond scans, while molecular $H_2O$, $H_2O_2$, $O_2$, and OH served as non-bonded references. To connect bulk properties to metal/oxide interface environments relevant to segregation simulations, the bare Al(111) surface and $Al_2O_3$ corundum surfaces were included alongside atomic oxygen adsorption at six high-symmetry sites (octahedral, tetrahedral, bridge, fcc-hollow, hcp-hollow, and on-top).

To parameterize the Al–Sn metallic chemistry and enable the convex-hull validation shown in Figure S1, four intermetallic phases were included: cubic $Al_3Sn$, tetragonal $Al_3Sn$, hexagonal AlSn, and cubic $Al_2Sn$. Each phase was incorporated through a formation-energy expression referenced to fcc Al and diamond-cubic α-Sn, together with compressed and expanded EOS variants (c1–c3, e1–e3 for cubic $Al_3Sn$; c1, e1 for the other three phases) to constrain the curvature of the energy landscape near equilibrium. Energy-difference entries between symmetry-distinct optimized variants and between compressed and expanded pairs were added to further constrain the relative stability of competing structures.

Following the development of the initial Al/Sn/O/H description, the force field was extended to the Sn–Al–O ternary oxide chemistry by incorporating three $AlSnO_3$ polymorphs (cubic, hexagonal, monoclinic) and the $Al_2Sn_2O_7$ phase as additional training targets. Each was included at its optimized geometry along with compressed and expanded variants (c1–c3, e1–e3) to capture the bulk response of the ternary oxide phases.

## Thermodynamic Stability: Sn-Al convex hull

The convex hull of formation energies for the Al–Sn system, depicted in Figure S1, provides a direct comparison between ReaxFF and DFT calculations. The ReaxFF results reproduce the Al–Sn convex hull well, capturing both the global curvature and the energetically favorable intermediate phases across the compositional range. The predicted formation energies closely follow the DFT trend, indicating that the force field reliably describes the thermodynamic stability of Al–Sn alloys. Key intermetallic phases, $Al_3Sn$ (tetragonal and cubic), $Al_2Sn$, and AlSn, are well represented, with only minor deviations in relative energies. A noticeable deviation is observed around the $Al_2Sn$ and AlSn compositions, where ReaxFF predicts a deeper energy minimum than DFT, suggesting a slight over-stabilization of these phases. A small shift in relative stability between closely competing $Al_3Sn$ structures is also evident. Such differences are consistent with the inherent approximations of reactive force fields, which are parameterized to reproduce a broad range of material properties rather than exact phase energetics. The overall agreement demonstrates that the parameterization captures the essential thermodynamic landscape of the Al-Sn system, supporting its use in large-scale simulations of alloy evolution, phase stability, and reactive processes.

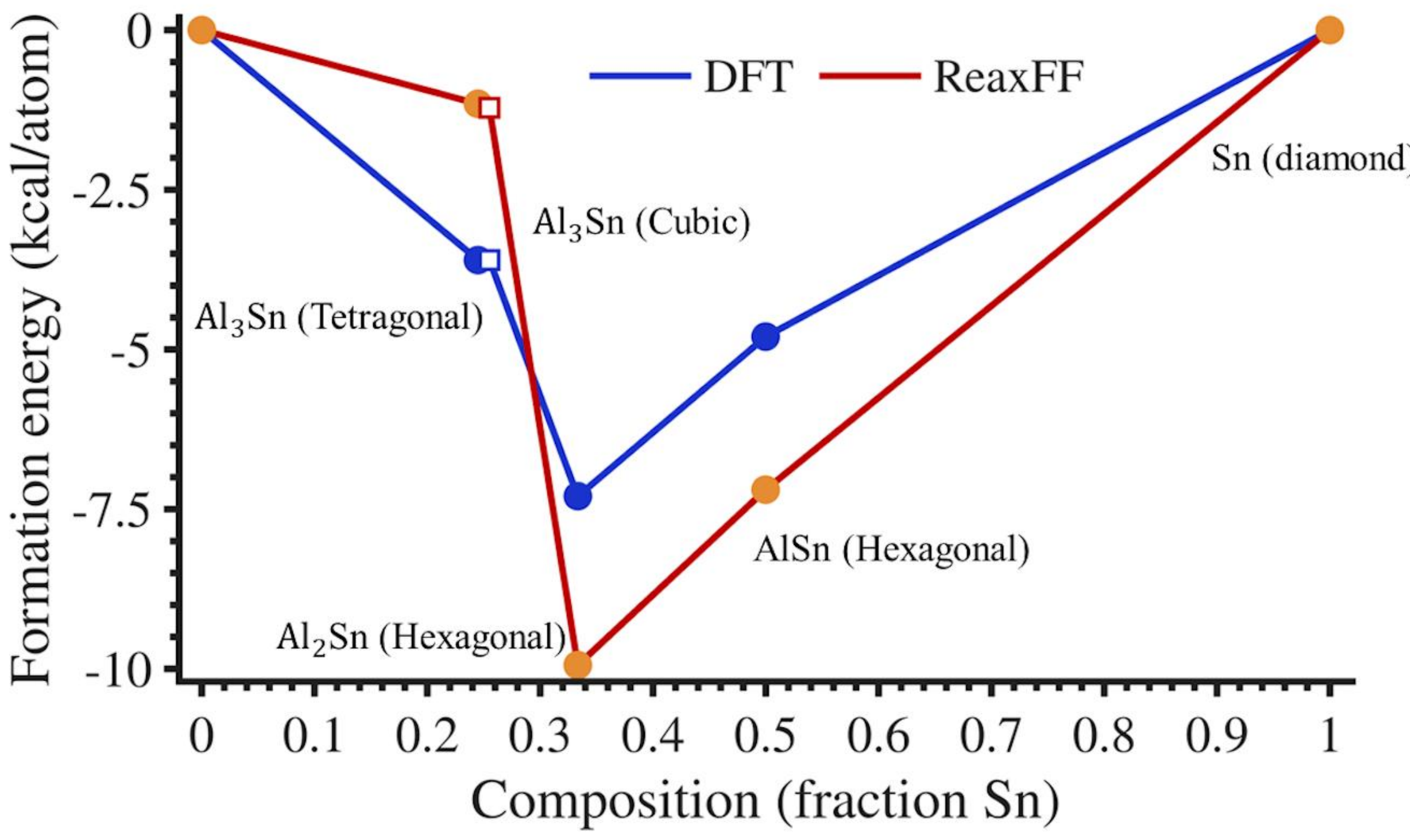

Figure S1. Convex hull of formation energy for Al–Sn alloy phases as a function of composition (Sn fraction). Formation energies from DFT (Materials Project) are compared with calculations from the ReaxFF potential. The stable/intermediate phases (cubic $Al_3Sn$, tetragonal $Al_3Sn$, cubic $Al_2Sn$, and hexagonal AlSn) are indicated along the convex hull.

## 1(b) Extension to Sn/Li/O/H: Sn-Li, Sn-Li-O, Sn-Li-H, and Sn-Li-O-H chemistry:

Starting from the Sn/Al/O/H parameters developed in Section 1(a), the force field was extended to describe Sn–Li chemistry in the presence of O and H. The training set for this extension covers the Li–Sn alloy phase space, Li–Sn–O ternary oxides, Li–H and Sn–H hydrides, and the bond-dissociation and elastic-distortion scans required to constrain the new Li-containing terms.

Five intermetallic phases spanning the full Li-Sn compositional range: $Li_{17}Sn_4$, $Li_7Sn_2$, $Li_3Sn$, $Li_5Sn_2$, and monoclinic LiSn, were included through formation-energy expressions referenced to bcc Li and diamond-cubic α-Sn. Additional distorted variants were added to enforce the correct curvature of the energy landscape near equilibrium, so that both the energetic ordering and the local shape of the energy wells are captured simultaneously. Systematic distortions were applied to orthorhombic $Li_7Sn_2$, generating energy-strain curves for the bulk (B) mode and for every symmetry-independent elastic mode ($C_{11}$, $C_{22}$, $C_{33}$, $C_{12}$, $C_{13}$, $C_{23}$, $C_{44}$, $C_{55}$, $C_{66}$), each sampled with six compressive (c1-c6) and six tensile (e1-e6) points.

Four Li-Sn-O compounds: $Li_2SnO_3$, $Li_8SnO_6$ (trigonal), $Li_2SnO_4$ (triclinic), and $Li_2SnO_6$ (monoclinic), were incorporated in their equilibrium forms, with $Li_2SnO_3$ additionally sampled in distorted configurations to capture variations in coordination and bonding. H–H dissociation was trained against a full 16-point $H_2$ scan, and Li–Li dissociation against a 15-point scan. The $SnH_x$ series (SnH, $SnH_2$, $SnH_3$, $SnH_4$) was included in molecular form, together with three $SnH_4$ solid polymorphs (monoclinic, orthorhombic, hexagonal). Lithium hydride was included as the rock-salt LiH phase with a systematic Li–H distance scan and a uniform-expansion series on the NaCl lattice. Equilibrium bond distances, bond angles, and Sn–Li cluster configurations (SnLi_cluster_HLi, SnLi_cluster_HSn, SnLi_cluster_H_SnLi) were included to sample the local bonding environments representative of segregated Sn–Li interfaces with O/H species, the target chemistry of the production simulations. LiOH polymorphs were added to anchor the Li–O–H corner of the phase space.

## Thermodynamic Stability: Sn-Li convex hull

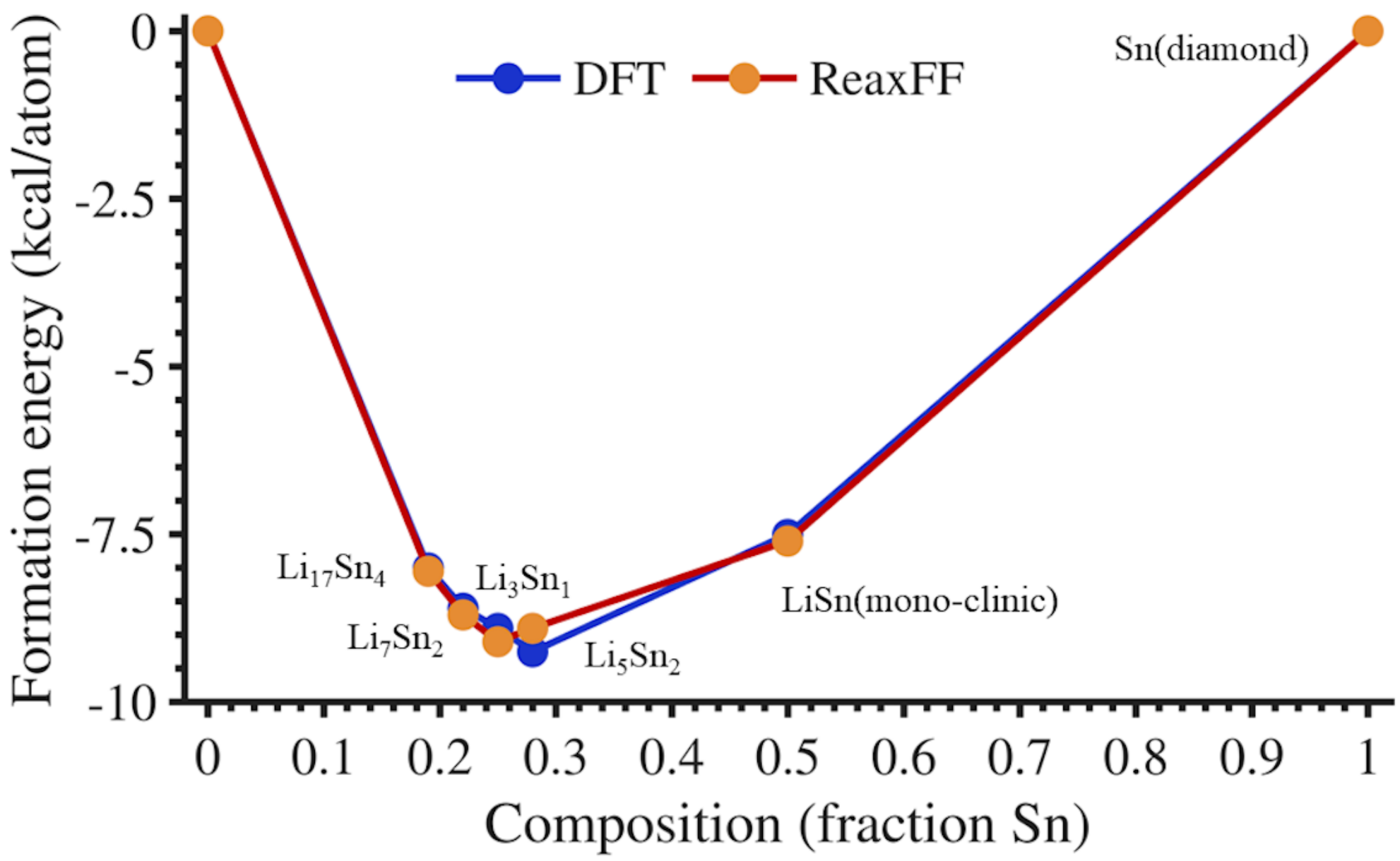

Figure S2. Convex hull of formation energy for Li–Sn alloy phases as a function of composition (fraction of Sn). Formation energies obtained from Density Functional Theory (DFT) are compared with calculations from the ReaxFF potential. The stable/intermediate phases ($Li_{17}Sn_4$, $Li_7Sn_2$, $Li_3Sn_1$, $Li_5Sn_2$, and LiSn in the monoclinic phase) are indicated along the convex hull.

The convex hull of formation energies for the Sn-Li system, depicted in Figure S2, provides a direct comparison between ReaxFF and DFT calculations. The ReaxFF results reproduce the Li–Sn convex hull well, capturing both the global curvature and the energetically favorable intermediate phases across the compositional range. The predicted formation energies closely follow the DFT trend, indicating that the force field reliably describes the thermodynamic stability of Li–Sn alloys. Overall, the intermetallic phases $Li_{17}Sn_4$, $Li_7Sn_2$, $Li_3Sn$, $Li_5Sn_2$, and monoclinic LiSn are all well represented. A slight shift in the preferred alloy structure is observed, where ReaxFF favors $Li_3Sn$ over $Li_5Sn_2$ relative to DFT, reflecting a small redistribution of stability among closely competing phases. The overall agreement demonstrates that the parameterization captures the essential thermodynamic landscape of the Li–Sn system, supporting its use in large-scale simulations of alloy evolution, phase stability, and reactive processes.

## Mechanical Response: Elastic Constants of $Li_7Sn_2$

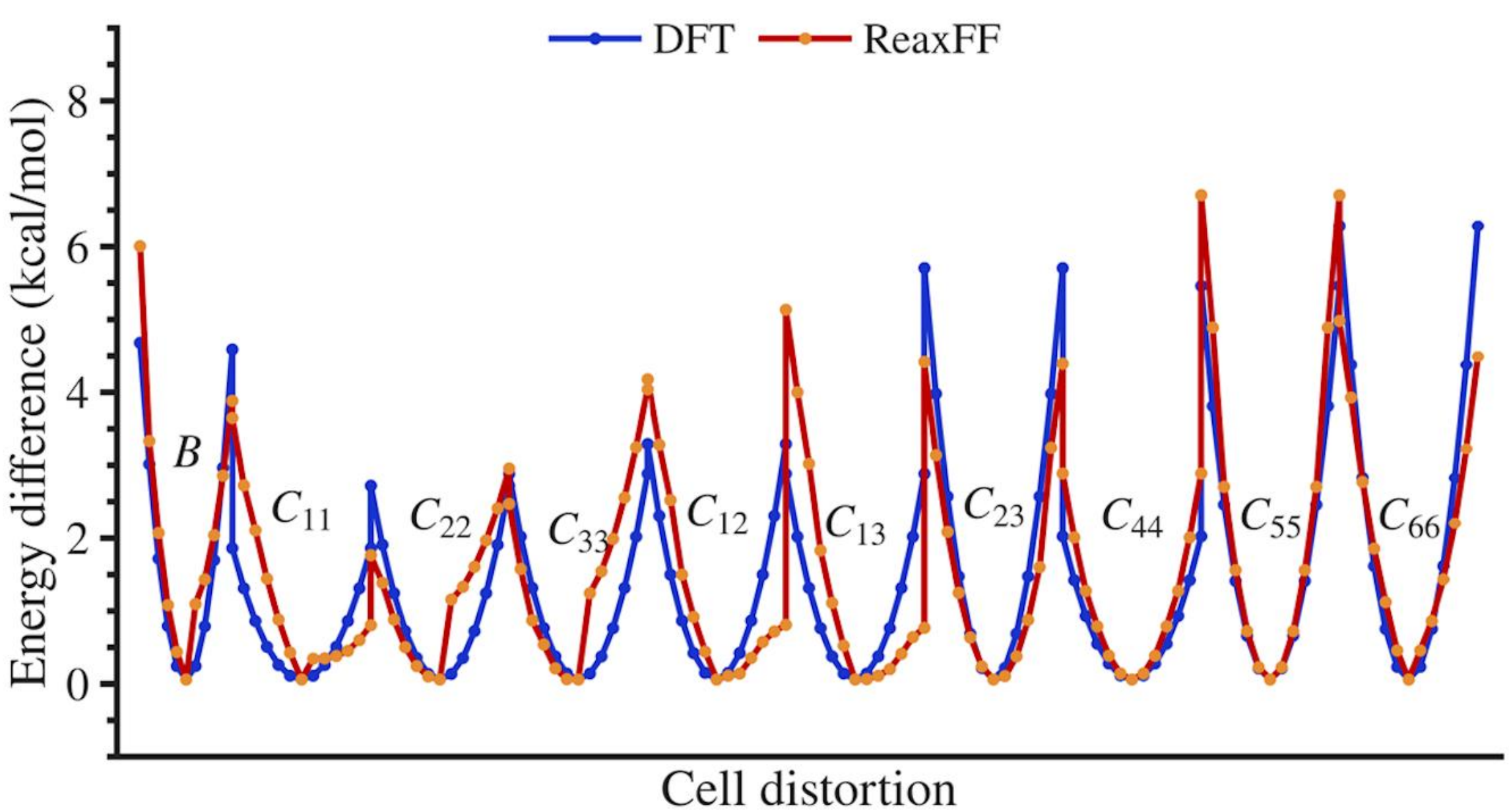

Figure S3. Elastic constant data for orthorhombic $Li_7Sn_2$. Energy differences (kcal $mol^{-1}$) are plotted as a function of cell distortion for bulk (B) and elastic modes ($C_{11}$–$C_{66}$), comparing ReaxFF (red) with DFT data (blue, Materials Project).

Figure S3 illustrates the energy variation as a function of the cell distortion used to probe the elastic constants of $Li_7Sn_2$ in its orthorhombic structure, comparing DFT and ReaxFF predictions. The distortion modes correspond to the bulk (B) and individual elastic constants ($C_{11}$, $C_{22}$, $C_{33}$, $C_{12}$, $C_{13}$, $C_{23}$, $C_{44}$, $C_{55}$, $C_{66}$). Overall, the ReaxFF results agree well with the DFT calculations, capturing both the curvature and the relative energy trends across all deformation modes. The potential reproduces the stiffness behavior and the symmetry of the energy wells with reasonable accuracy, indicating its reliability for mechanical-response predictions. The main deviation is observed for the $C_{13}$ distortion, where ReaxFF slightly overestimates the energy compared with DFT, producing a modestly stiffer response than the reference. This discrepancy is confined to a single mode and does not propagate into the diagonal elastic constants that dominate the mechanical response. The overall consistency confirms that the developed force field provides a robust representation of the elastic behavior of $Li_7Sn_2$.

## Oxide and Hydride Energies

The formation energies predicted by ReaxFF for the Sn–Al–O, Sn–Li–O, LiH, and $SnH_4$ systems are compared with DFT values from the Materials Project[2] in Table S1. For the Sn–Al–O structures, ReaxFF captures the relative energetic ordering of the different polymorphs with absolute deviations of 24–59 kcal mol$^{-1}$ (9–16 % in relative terms), preserving the trend across compositions and crystal structures. For the Sn–Li–O structures, agreement is excellent for the simpler phases $Li_2SnO_3$ and $Li_2SnO_4$ (deviations of 2.6 and 3.6 kcal mol$^{-1}$, respectively, i.e. ≈1 %). Larger deviations appear for $Li_2SnO_6$ (27 kcal mol$^{-1}$, 13 %) and $Li_8SnO_6$ (125 kcal mol$^{-1}$, 17 %), reflecting the well-known difficulty of capturing the energetics of highly coordinated, multi-component ionic environments within a single reactive parameter set.

For the LiH rock-salt structure, ReaxFF predicts a more negative formation energy than DFT (10 kcal mol$^{-1}$ deviation), indicating a slight over-binding of the Li–H interaction; this reflects the difficulty of representing ionic Li–H bonding alongside covalent Sn–H and metallic Li–Sn interactions within a unified parameter set. The $SnH_4$ polymorphs show mixed agreement: the monoclinic phase is reproduced almost exactly, and the orthorhombic (I) phase shows only a small deviation (3.7 kcal mol$^{-1}$), whereas the orthorhombic (II) and hexagonal polymorphs show substantial residuals. The hexagonal polymorph in particular lies far above the DFT reference; this is the least well-described environment in the set and is noted here as a known limitation of the current parameter set. Because the Sn–Li segregation simulations for which this force field is developed do not sample condensed $SnH_4$ polymorph configurations, this deviation is not expected to affect the production results, but it delimits the applicability of the potential and would require further refinement before use in hydride-specific studies.

Table S1: Comparison of formation energies (kcal/mol) for Sn-based oxides, hydrides, and Li-containing compounds: ReaxFF vs DFT (Materials Project)

| Structures | ReaxFF (kcal/mol) | DFT (kcal/mol) (Materials Project) |
|---|---|---|
| | **Sn-Al-O Structures** | |
| **$Sn_2Al_2O_7$** | -605.74 | -546.70 |
| **$SnAlO_3$ (Hexagonal)** | -292.83 | -257.80 |
| **$SnAlO_3$ (Cubic)** | -130.52 | -155.60 |
| **$SnAlO_3$ (Monoclinic)** | -304.38 | -280.26 |
| | **Sn-Li-O Structures** | |
| **$Li_2SnO_3$ (Monoclinic)** | -309.71 | -307.08 |
| **$Li_8SnO_6$ (Trigonal)** | -864.61 | -739.20 |
| **$Li_2SnO_4$ (Triclinic)** | -291.06 | -294.63 |
| **$Li_2SnO_6$ (Monoclinic)** | -230.08 | -203.40 |
| | **Li-H structure** | |
| **Cubic, Rocksalt** | -32.64 | -22.56 |
| | **$SnH_4$ polymorphs structures** | |
| **Monoclinic** | 13.68 | 13.69 |
| **Orthogonal (I)** | 41.14 | 44.82 |
| **Orthogonal (II)** | -3.36 | 22.20 |
| **Hexagonal** | 171.79 | 60.15 |

Overall, the ReaxFF potential provides a consistent and transferable description of the Sn/Al/Li/H/O chemistry, capturing the elemental Al and Sn reference states, the Al-O-H oxide and hydroxide chemistry, together with Al(111) oxygen adsorption. The convex-hull thermodynamics of the Al–Sn and Li–Sn intermetallics (Figures S1 and S2), the elastic response of $Li_7Sn_2$ (Figure S3), and the formation energies of the ternary oxides and hydrides (Table S1), with residual deviations confined to the structurally most complex or compositionally most demanding environments, further confirm the adequacy of the force field developed for large-scale atomistic simulations of Sn–Li segregation in the presence of O and H species.